\documentclass[aps,prb,twocolumn,groupedaddress,showpacs,floatfix,letterpaper]{revtex4}

\usepackage{graphicx}

\newcommand{\re}{{\bf r}  }
\newcommand{\rp}{{\bf r'}  }

\begin{document}

%\preprint{}

\title{Optimized Effective Potential made simple: 
Orbital functionals, orbital shifts, and the exact Kohn-Sham exchange
potential}

\author{Stephan K\"ummel}
%\email[]{skuemmel@tulane.edu}
%\homepage[]{http://loki.phy.tulane.edu/~kuemmel/}
\affiliation{Department of Physics and Quantum Theory Group, Tulane
  University, New Orleans, Louisiana 70118, USA}

\author{John P.\  Perdew}
%\email[]{perdew@tulane.edu}
%\homepage[]{}
\affiliation{Department of Physics and Quantum Theory Group, Tulane
  University, New Orleans, Louisiana 70118, USA}

\date{\today}
%submitted December 7 2002, accepted (straightforward) march 2003

\begin{abstract}

The optimized effective potential (OEP) is the exact Kohn-Sham
potential for explicitly orbital-dependent energy functionals, e.g.,
the exact exchange energy. We give a proof for the OEP equation which
does not depend on the chain rule for functional derivatives and
directly yields the equation in its simplest form: a certain
first-order density shift must vanish. This condition explains why the
highest-occupied orbital energies of Hartree-Fock and exact-exchange
OEP are so close. More importantly, we show that the
exact OEP can be constructed iteratively from the first-order shifts
of the Kohn-Sham orbitals, and that these can be calculated
easily. The exact exchange potential $v_\mathrm{x}(\re)$ for spherical
atoms and three-dimensional sodium clusters is calculated. Its
long-range asymptotic behavior is investigated, including the approach
of $v_\mathrm{x}(\re)$ to a non-vanishing constant in particular
spatial directions. We calculate total and orbital energies and static
electric dipole polarizabilities for the sodium clusters employing the
exact exchange functional. Exact OEP results are compared to the
Krieger-Li-Iafrate (KLI) and local density approximations.

\end{abstract}

\pacs{71.15.Mb,31.15.Ew,36.40.-c,73.22.-f 
\hfill Physical Review B
}
% 71.15.Mb density functional theory (solids), PRB section B15
% 31.15.Ew  Density functional theory in atomic and molecular physics
% 31.15.-p calculations and mathematical techniques (atoms and
%             molecules) 
% 36.40.-c Atomic and molecular clusters
% 73.22.-f Electronic structure of nanoscale materials: clusters, 
%          nanoparticles, nanotubes, and nanocrystals

\maketitle

\section{Introduction}

Density functional theory (DFT) in general and the Kohn-Sham scheme in
particular are among the most important tools for electronic structure
calculations. This success is based on a rigorous foundation in
terms of the Hohenberg-Kohn theorem on the one hand \cite{hk}, and
increasingly sophisticated approximations to the exchange-correlation
functional $E_\mathrm{xc}$ on the other. The local density
approximation (LDA) \cite{ks} proved to be of far greater
applicability than originally expected. Its great advantages are
simplicity and reliability in the sense that its shortcomings are
qualitatively predictable, but the price to be paid is limited
accuracy. Generalized gradient approximations (GGA's)
\cite{realcut,becke88,lyp,pbe} improved accuracy to a level that made
DFT useful for many chemical applications. But despite their
many successes, GGA's cannot be regarded as the final stage of
functional development \cite{jaclad}. Further improvements in accuracy
are expected from functionals that include partial or full exact
exchange. Global \cite{beckehy1,jphy,ernzerhofhy,burkehy,scuseriahy}
and local \cite{scuserialhy} hybrids and hyper-GGA's
\cite{jaclad,newhgga} fall into this class.

Using the exact exchange energy in DFT calculations is promising from
many points of view. Full exact exchange cancels the spurious Hartree
self-interaction energy, curing in a systematic way one of the most
notorious DFT problems. It leads to the correct high-density limit
\cite{jpreview}, which is exchange dominated. Great improvements in
the Kohn-Sham eigenvalue spectrum \cite{goerasymp,goni}, semiconductor
bandstructure and excitations
\cite{bylander,staedeleprl,staedeleprb,kimexcitons} and nonlinear
optical properties \cite{chains} have been reported using the exact
exchange energy. The exact exchange functional also leads to the
correct asymptotic $(r\rightarrow\infty)$ behavior of the Kohn-Sham
potential, which turns 
out to show surprising features not present in any of the common
approximations, as discussed in Refs.\ \onlinecite{goerasymp,oep1} and
below. Finally, the idea of incorporating one more known exact
ingredient into the energy functional is appealing in its own
right. 

But the goal of using the exact exchange energy or, more generally,
any orbital functional self-consistently in Kohn-Sham calculations
raises the question of how to construct the corresponding
exchange-correlation potential. $E_\mathrm{xc}$ is still implicitly a
functional of the density, but explicitly only its dependence on the
set of occupied Kohn-Sham orbitals is known. Thus,
calculating the corresponding exchange-correlation potential
$v_\mathrm{xc}(\re)=\delta E_\mathrm{xc}[n]/\delta n(\re)$ cannot be
done straightforwardly by explicitly taking the functional derivative
with respect to the density $n(\re)$. Instead, the potential must be
calculated from the optimized effective potential (OEP) integral
equation \cite{sharp,talman,sahni}. This approach is size-consistent
if the orbital functional is invariant under unitary transformations
of the occupied orbitals and is not too radically nonlocal
\cite{kp2}. Size consistency is achieved for exact exchange, hybrid
functionals
\cite{beckehy1,jphy,ernzerhofhy,burkehy,scuseriahy,scuserialhy},
meta-GGA's \cite{pkzb,tp} and hyper-GGA's \cite{jaclad,newhgga}.

Due to its complexity, direct solutions of the OEP integral equation
so far have been restricted to effectively one-dimensional systems
with spherical symmetry
\cite{talman,kli2,engel1,engel2,kotani1,kotani2,grabo1,graborev}, and
the calculations for finite systems have been based on the program
developed in Ref.\ \onlinecite{talman}. Hopes are low to generalize
this approach to higher dimensions. The OEP for three-dimensional
systems has been constructed by directly evaluating the response
function using basis sets
\cite{staedeleprl,staedeleprb,goerlev,goer99,bartlett}. This approach
has proven successful, but requires considerable technical expertise:
Summing over not only occupied but also unoccupied Kohn-Sham orbitals
is required, and the necessary inversion of the response function can
be cumbersome. And while the OEP total energy can be accurately
obtained, the potentials and Kohn-Sham eigenvalues for finite systems
may suffer from basis-set limitations. A detailed discussion of the
method and its limitations can be found in Refs.\
\onlinecite{hirata,lhf,hamel}. Alternatively, the short-range part of
the potential itself has been expanded in a basis set and the
expansion coefficients chosen to minimize the total energy
\cite{fritsche,colle,yangwu}. This approach is appealing because of
its directness. But the $-e^2/r$-decay that the OEP is thereby forced
to take everywhere in the asymptotic region of a finite system has to
be reconsidered in the light of new conclusions about the asymptotic
behavior of $v_\mathrm{x}(\re)$ presented in Refs.\
\onlinecite{goerasymp,oep1} and discussed below. Also, the method is
hard to implement within fully numerical schemes for electronic
structure calculations, and those continue to grow in importance
\cite{pgrev,kronik,beck}.

In this manuscript we discuss how the OEP can easily be constructed
from the first-order orbital shifts that have been introduced into OEP
theory to justify the KLI approximation \cite{kli2}. We first present
in Section \ref{secdensityshift} a new proof for the OEP equation that
explains why this complicated equation can be cast in a simple
form. In Section \ref{sechf} we compare and contrast Hartree-Fock
theory and exact-exchange OEP. Different methods to construct the OEP
from the orbital shifts are analyzed in Section \ref{secconstruction},
together with indicators for the accuracy of an OEP solution. The
exact-exchange OEP for spherical atoms and three-dimensional sodium
clusters is calculated. Its long-range asymptotic behavior is
investigated in Section \ref{secasymptotics}. We show that the
behavior $\lim_{r\rightarrow\infty}v_\mathrm{x}(\re)=-e^2/r+C$, with
$C \neq 0$ on nodal surfaces of the highest occupied Kohn-Sham
orbital, is a common situation for metal clusters. Finally, in Section
\ref{secpol} we calculate the static electric polarizability for the
neutral, even-electron clusters $\mathrm{Na}_2$ --
$\mathrm{Na}_{10}$ in the exchange-only approximation, and compare
LDA, KLI and exact OEP results.

\section{The OEP equation in terms of the first-order density shift}  
\label{secdensityshift}

As shown by Krieger, Li and Iafrate \cite{kli2} and further clarified
by Grabo, Kreibich, Kurth and Gross \cite{graborev}, the OEP integral
equation for the spin-dependent exact Kohn-Sham exchange-correlation
potential $v_\mathrm{xc\sigma}(\re)$ can be written in a form that takes a
very simple interpretation. At the end of a long argument involving
repeated application of the chain rule for functional derivatives and
linear response theory, the OEP equation is written in the form
\begin{equation}
\label{oep}
\sum_{i=1}^{N_\sigma} \psi_{i \sigma}^*(\re) \varphi_{i \sigma}(\re) +
c.c. 
=0.
\end{equation}
Eq.\ (\ref{oep}) says that the optimum (i.e., yielding the lowest
Kohn-Sham energy) potential $v_\mathrm{xc\sigma}(\re)$ to replace the
orbital-dependent potential 
\begin{equation}
\label{defu}
u_{\mathrm{xc}i\sigma}(\re)=\frac{1}{\varphi_{i \sigma}^*(\re)}
 \frac{\delta E_\mathrm{xc}\left[\left\{ \varphi
                                 \right\}
                          \right]}
      {\delta \varphi_{i \sigma}(\re)}
\end{equation}
is the one that makes the change in the density vanish to first order
in the perturbation 
\begin{equation}
\label{pertv}
\Delta v_{i\sigma}(\re)=u_{\mathrm{xc}i\sigma}(\re)-v_\mathrm{xc\sigma}(\re),
\end{equation}
when this perturbation is applied to the Kohn-Sham system. The
$\varphi_{i \sigma}(\re)$ in Eq.\ (\ref{oep}) are the Kohn-Sham
orbitals, i.e., the solutions of
\begin{equation}
\label{kseq}
\left(\hat{h}_{\mathrm{KS}\sigma}-\varepsilon_{i \sigma}\right)
 \varphi_{i \sigma}(\re)= 0,
\end{equation}
where
$\hat{h}_{\mathrm{KS}\sigma}=-(\hbar^2/2m) \nabla^2
 +v_{\mathrm{KS}\sigma}(\re)$ is the Kohn-Sham Hamiltonian. The
self-consistent Kohn-Sham potential   
\begin{equation}
\label{kspot}
v_{\mathrm{KS}\sigma}(\re)=v(\re)+v_\mathrm{H}(\re)+
v_{\mathrm{xc}\sigma}(\re)
\end{equation}
is the sum of the external potential $v(\re)$,
the Hartree potential $v_\mathrm{H}(\re)=\int d^3 r' e^2 n(\rp)/|\re -\rp| $,
and the spin-dependent exchange-correlation potential
$v_{\mathrm{xc}\sigma}(\re)=\delta E_\mathrm{xc}[n]/\delta
n_\sigma(\re)$. ($v(\re)$ can also depend upon $\sigma$, but usually
does not.)
$\psi_{i \sigma}(\re)$ is the negative of the first-order
perturbation-theory shift that results if $\varphi_{i\sigma}(\re)$
is subjected to the perturbation of Eq.\ (\ref{pertv}), i.e,
\begin{eqnarray}
\label{psipert}
\lefteqn{
-\psi_{i \sigma}^*(\re)=}
\\ & &
\sum_{\stackrel{\scriptstyle j=1}{j\neq i}}^{\infty}
\frac{\int \varphi_{i \sigma}^*(\rp)
  \left[ u_{\mathrm{xc}i\sigma}(\rp) - v_{\mathrm{xc}\sigma}(\rp) \right]
  \varphi_{j \sigma}(\rp) \, \mathrm{d}^3 r'}
{\varepsilon_{i \sigma}-\varepsilon_{j \sigma} }
\varphi_{j \sigma}^*(\re). \nonumber 
\end{eqnarray}
For the sake of simplicity, we here assume non-degenerate orbitals. The
extension to the degenerate case is unproblematic: As shown in
Appendix B of Ref.\ \onlinecite{kli1}, the restriction $j\neq i$ in
the sum is to be replaced by $\varepsilon_{j\sigma} \neq
\varepsilon_{i\sigma}$. 

Note that, if the exact exchange energy
\begin{equation}
\label{exx}
E_\mathrm{x}[\{\varphi\}]=
-\frac{e^2}{2} 
\sum_{\stackrel{\scriptstyle i,j=1}{\sigma=\uparrow,\downarrow}}^{N_\sigma}
\int \int 
\frac{
 \varphi_{i \sigma}^*(\re) \varphi_{j \sigma}^*(\rp)
 \varphi_{j \sigma}(\re) \varphi_{i \sigma}(\rp) 
 }{
 \left| \re-\rp \right| 
}
\mathrm{d}^3r' \, \mathrm{d}^3r,
\end{equation}
is substituted for
$E_\mathrm{xc}$, then Eq.\ (\ref{defu}) just yields the
orbital-dependent Hartree-Fock potential\cite{gl95}
\begin{equation}
u_{\mathrm{x}i\sigma}(\re)=-\frac{e^2}{\varphi_{i\sigma}^*(\re)}
\sum_{j=1}^{N_\sigma} \varphi_{j\sigma}^*(\re) \int 
\frac{\varphi_{i\sigma}^*(\rp)\varphi_{j\sigma}(\rp)}{|\rp-\re|}
\, \mathrm{d}^3 r'.
\end{equation}

Eq.\ (\ref{oep}) is intuitively appealing for the following reasons:
(1) It is a statement about the electron spin density, the central
quantity of density functional theory. Going over from the optimized
effective potential to the orbital-dependent potentials should not
change the density much. In important limits (one- and two-electron
ground states, and the uniform electron gas), the density should not
change at all. (2) Once the OEP orbitals and orbital energies are
fixed, Eq.\ (\ref{oep}) is homogeneous of degree one in the
perturbation of Eq.\ (\ref{pertv}), making
$v_{\mathrm{xc}\sigma}(\re)$ a kind of average of the
$u_{\mathrm{xc}i\sigma}(\re)$. A simple average was proposed in Eq.\
(75) of Ref.\ \onlinecite{pzsic}, and averaged potentials are popular
in time-dependent DFT \cite{legrand}. A sophisticated but still
approximate average was proposed in Ref.\ \onlinecite{kli2}. (3) Eq.\
(\ref{oep}) contains just enough information to define
$v_{\mathrm{xc}\sigma}(\re)$ uniquely to within an additive constant,
given the OEP orbitals and orbital energy differences: Imagine
discretizing the $\re$ space into a collection of $M$ points $\re_1,
... ,\re_M$, and solving Eq.\ (\ref{oep}) as a set of $M-1$ linear
equations for the $M-1$ unknowns $v_{\mathrm{xc}\sigma}(\re_1),
... ,v_{\mathrm{xc}\sigma}(\re_{M-1})$. Because the number of electrons
cannot change, only $M-1$ equations are linearly independent, leaving
$v_{\mathrm{xc}\sigma}(\re_M)$ as a free parameter. This parameter is
normally chosen so that $\lim_{r\rightarrow \infty}
v_{\mathrm{xc}\sigma}(\re)=0$.

The OEP equation has previously been investigated from different perspectives
\cite{sharp,talman,goerlev,kli2,graborev,gl95,shaginyan,engdreiz,newlevy}. But 
an equation as charmingly simple as Eq.\ (\ref{oep}) also deserves a simple
proof. We provide one in the following. For notational simplicity, we
drop the spin index $\sigma$ in the rest of this section. 

We start from a given expression for the total energy in terms of some set of
orbitals $\{\phi\}$,
\begin{eqnarray}
\lefteqn{ \hspace{-8mm}
E_v^\alpha[\{\phi\}]=
 \frac{\hbar^2}{2m}\sum_{i=1}^N\int  |\nabla \phi_i(\re)|^2 \, \mathrm{d}^3r
+
 \int v(\re) n(\re) \, \mathrm{d}^3r
}
\nonumber \\&&
+
 \frac{e^2}{2} \int \int \frac{n(\re) n(\rp)}{|\re-\rp|} 
 \mathrm{d}^3r' \mathrm{d}^3r
+ \alpha E_\mathrm{xc}[\{\phi\}],
\label{ephi}
\end{eqnarray}
where
\begin{equation}
\label{defn}
n(\re)=\sum_{i=1}^N |\phi_i(\re)|^2.
\end{equation}
The real parameter $\alpha$ is introduced for the sake of later
arguments and can be set to 1 at the end. Based on Eq.\ (\ref{ephi}),
two different density functionals can be defined. The Kohn-Sham (i.e, OEP)
functional is defined by choosing the $\{\phi\}$ to be the Kohn-Sham
orbitals $\{\varphi\}$. This means that for a given density $n(\re)$, we
find a local potential $v_s(\re)$ such that the solutions of the
independent-particle Schr\"odinger equation 
\begin{equation}
\label{spseq}
\left(-\frac{\hbar^2}{2m}+v_s(\re)\right)\varphi_i(\re)=
\varepsilon_i \varphi_i(\re)
\end{equation} 
yield the given density via Eq.\ (\ref{defn}). The orbitals are clearly
functionals of $v_s(\re)$ which is itself a functional of $n(\re)$. The
Kohn-Sham energy functional is then
\begin{equation}
\label{eks}
E_v^{\mathrm{OEP},\alpha}[n]=E_v^\alpha[\{\varphi[n]\}].
\end{equation} 
Eq.\ (\ref{eks}) is minimized by the density 
$n^{\mathrm{OEP},\alpha}(\re)$, and this density is obtained by
choosing for $v_s(\re)$ the potential of Eq.\ (\ref{kspot}) (with
$v_\mathrm{xc}$ multiplied by $\alpha$, see below).  We define a
second, ``orbital'' density functional
\begin{equation}
\label{eorb}
E_v^{\mathrm{orb},\alpha}[n]=
E_v^\alpha[\{\varphi^{\mathrm{orb},\alpha}[n]\}]
\end{equation}
as the energy obtained by evaluating Eq.\ (\ref{ephi}) with a
different set of orbitals $\{\varphi^{\mathrm{orb},\alpha}\}$. The
$\varphi_i^{\mathrm{orb},\alpha}$ are defined as those normalized
linearly-independent orbitals that yield a given density via Eq.\
(\ref{defn}) and deliver the lowest extremum of
$E_v^\alpha[\{\phi\}]$. (This definition is discussed in Appendix
\ref{appendixa}). The density $n^{\mathrm{orb},\alpha}(\re)$ that
minimizes $E_v^{\mathrm{orb},\alpha}[n]$ results from the orbitals
that are solutions of
\begin{eqnarray}
\left(-\frac{\hbar^2}{2m}\nabla^2 + v(\re) + v_\mathrm{H}(\re) +
{u_{\mathrm{xc}i}^\alpha}^*(\re)\right) \varphi_i^{\mathrm{orb},\alpha}(\re)
&=&
\nonumber \\ 
\varepsilon_i^{\mathrm{orb},\alpha}\varphi_i^{\mathrm{orb},\alpha}(\re).
\label{orbeq}
&&
\end{eqnarray}
The requirement of a local, orbital-independent potential in the definition of
$E_v^{\mathrm{OEP},\alpha}[n]$ can be understood as an additional
constraint not present in $E_v^{\mathrm{orb},\alpha}[n]$. Therefore,
if we define a functional $A^\alpha[n]$ by
%$E_v^{\mathrm{orb},\alpha}[n] \leq E_v^{\mathrm{OEP},\alpha}[n]$.
\begin{equation}
\label{defa}
E_v^{\mathrm{orb},\alpha}[n] = E_v^{\mathrm{OEP},\alpha}[n] + A^\alpha[n],
\end{equation}
we know that $A^\alpha[n]\leq 0$. From the minimum principle for
$E_v^{\mathrm{orb},\alpha}[n]$, and from the preceding statement, it
is clear that
\begin{equation}
\label{ineq}
E_v^{\mathrm{orb},\alpha}[n^{\mathrm{orb},\alpha}]<
E_v^{\mathrm{orb},\alpha}[n^{\mathrm{OEP},\alpha}]<
E_v^{\mathrm{OEP},\alpha}[n^{\mathrm{OEP},\alpha}].
\end{equation}
For a given orbital functional, one achieves a lower energy with the
orbital-dependent potentials than with the OEP.

Eq.\ (\ref{oep}) can now be deduced directly from the
$\alpha$-dependence of the two energy functionals. We first observe
that
\begin{equation}
\label{vxca}
v_\mathrm{xc}^{\mathrm{OEP},\alpha}([n];\re)
=
\alpha \frac{\delta E_\mathrm{xc}^\mathrm{OEP}[n]}{\delta n(\re)},
\end{equation}
and
\begin{equation}
\label{ua}
u_{\mathrm{xc}i}^\alpha(\{\phi\};\re)
=
\alpha \frac{1}{\phi_i^*(\re)}
 \frac{\delta E_\mathrm{xc}\left[\left\{ \phi
                                 \right\}
                          \right]}
      {\delta \phi_i(\re)},
\end{equation}
i.e., both the orbital-independent and the orbital-dependent potential
depend linearly on $\alpha$. We further observe that, in contrast to the
$\varphi_i^{\mathrm{orb},\alpha}(\re)$, the Kohn-Sham orbitals for a
given density do not
depend on $\alpha$: The Hohenberg-Kohn theorem tells us that, for a given
density, $v_s(\re)$ and thus the orbitals $\varphi_i(\re)$ are 
uniquely fixed. Finally, we observe that for $\alpha=0$, the
functionals Eq.\ (\ref{eks}) and Eq.\ (\ref{eorb}) are trivially
identical. (Thus, in Eq.\ (\ref{eorbpower}) below,
$E_v^{\mathrm{orb},0}[n]=E_v^{\mathrm{OEP},0}[n]$.) 

From the second observation and Eq.\ (\ref{ephi}) one deduces that, for
fixed external potential $v(\re)$ and fixed density $n(\re)$,
the $\alpha$-dependence of the OEP-functional is
\begin{equation}
\label{ekspower}
E_v^{\mathrm{OEP},\alpha}[n]=E_v^{\mathrm{OEP},0}[n]+
\alpha E_\mathrm{xc}[\{\varphi[n]\}].
\end{equation}
In contrast, $E_v^{\mathrm{orb},\alpha}[n]$ in general will not have
such a simple dependence on $\alpha$ since the
$\varphi_i^{\mathrm{orb},\alpha}$ for a given density will change with
$\alpha$. Thus, the power series
\begin{equation}
\label{eorbpower}
E_v^{\mathrm{orb},\alpha}[n]=E_v^{\mathrm{orb},0}[n] + 
\alpha E_{v,1}^{\mathrm{orb}}[n] +
\alpha^2 E_{v,2}^{\mathrm{orb}}[n] +
...,
\end{equation}
where formally 
$E_{v,1}^{\mathrm{orb}}[n]=
(\partial E_v^{\mathrm{orb},\alpha}[n]/\partial \alpha)|_{\alpha=0}$,
etc., will have non-vanishing terms beyond the one linear in
$\alpha$. However, the crucial fact to note is that the term linear in
$\alpha$ is the same in $E_v^{\mathrm{OEP},\alpha}[n]$ and
$E_v^{\mathrm{orb},\alpha}[n]$: At $\alpha=0$ (Hartree approximation),
the orbitals are the same, i.e.,
$\varphi_i^{\mathrm{orb},0}(\re)=\varphi_i(\re)$, and we know that
there can be no first-order change in $E_v^{\mathrm{orb},\alpha}[n]$
due to a change in the orbitals, because we defined the
$\varphi_i^{\mathrm{orb},\alpha}(\re)$ as extremizing orbitals. Therefore,
$E_{v,1}^{\mathrm{orb}}[n]=E_\mathrm{xc}[\{\varphi[n]\}]$. 

This means that the functional $A^\alpha[n]$ of Eq.\ (\ref{defa})
vanishes to linear order in $\alpha$. It is then intuitively clear that
the densities that minimize  $E_v^{\mathrm{orb},\alpha}[n]$ and
$E_v^{\mathrm{OEP},\alpha}[n]$ must also be the same to linear order in
$\alpha$. In other words, the difference to first order in $\alpha$ between
$n^{\mathrm{orb},\alpha}(\re)$ and $n^{\mathrm{OEP},\alpha}(\re)$ must
vanish, and, because of Eqs.\ (\ref{vxca}) and (\ref{ua}), this difference
is just given by the left-hand side of Eq.\ (\ref{oep}). A detailed version of
these arguments is presented in Appendix \ref{appendixb}. 

Note that the determinant of orbitals satisfying Eqs.\ (\ref{spseq}) or
(\ref{orbeq}) and achieving the lowest total energy need not necessarily
satisfy the Aufbau principle, but might have unoccupied orbitals with orbital
energies lying below occupied ones. Note further that, in the absence of a
magnetic field, all the orbitals can be chosen to be real.

\section{Hartree-Fock vs.\ Exchange-Only OEP}
\label{sechf}

As an important example of the ideas of Section \ref{secdensityshift},
let $E_{\mathrm{xc}}[\{\phi\}]$ in Eq.\ (\ref{ephi}) be the Fock
exchange integral of the occupied orbitals of Eq.\ (\ref{exx}). Then
the lowest extremum over linearly-independent orbitals in the definition
of $E_v^{\mathrm{orb},\alpha}[n]$ becomes the
minimum over orthonormal orbitals. For given
external potential $v(\re)$ and electron number $N$, the
$\varphi_i^{\mathrm{orb},\alpha=1}(\re)$ are the canonical 
Hartree-Fock orbitals. $E_v^{\mathrm{orb},\alpha=1}[n]$ is the density
functional for the Hartree-Fock energy defined by the Levy constrained
search \cite{consearch,dreizlergross} (but not known as an
explicitly calculable functional of the Kohn-Sham orbitals or of the
density). Its minimizing density $n^{\mathrm{orb},\alpha=1}(\re)$ is
the Hartree-Fock density, and
$E_v^{\mathrm{orb},\alpha=1}[n^{\mathrm{orb},\alpha=1}]$ is the
Hartree-Fock energy. 
The minimizing problem for this functional can be formulated within 
%The same energy and density can be obtained within
Kohn-Sham theory, but the resulting Kohn-Sham orbitals
$\varphi_i([n^{\mathrm{orb},\alpha=1}];\re)$ are {\it not} the
Hartree-Fock orbitals, although both sets of orbitals yield the same
density. The relation between Hartree-Fock theory and DFT has also been
studied in Ref.\ \onlinecite{holas}.

On the other hand, $E_v^{\mathrm{OEP},\alpha}[n]$ is by definition
\cite{sahni} the exact exchange-only density functional of Kohn-Sham
theory (an explicitly calculable functional of the Kohn-Sham
orbitals, and one which defines the lower limit of the integrand in the
coupling constant or adiabatic connection formula for $E_\mathrm{xc}[n]$
\cite{lp,gl}). The inequality following Eq.\ (\ref{defa}) shows that the
Hartree-Fock exchange energy for a given density includes a negative
contribution which from the exchange-only OEP perspective is a small
part of the correlation energy \cite{hfecsahni,hfecgross}. (The local
density approximation to this part of the correlation energy vanishes,
while its second-order gradient coefficient diverges \cite{sahni},
making it very hard to approximate.) Since $e^2$ plays the same
multiplicative role as $\alpha$, the exact exchange energy of
Kohn-Sham theory \cite{sahni} for a given density is purely of order
$e^2$, while the Hartree-Fock exchange energy for the same density
agrees with it to order $e^2$, differing in order\cite{sahni,lm}
$e^4$. Eq.\ (\ref{ineq}) also shows that the Hartree-Fock 
total energy must be less than the total energy of the exact
exchange-only OEP theory. These exact-exchange conclusions are
not new \cite{sahni}, yet 
%and were known to the authors of Ref.\ \onlinecite{sahni}, yet
these subtle distinctions continue to cause some confusion.

The Hartree-Fock and exchange-only OEP minimizing densities for a
given $v(\re)$ also agree to order $e^2$, differing from one another
and from the correlated ground-state density in order $e^4$. Since the
highest-occupied orbital energy controls the asymptotic decay of the
density, the highest occupied orbital energies must also agree to
order $e^2$ with one another, and with minus the correlated first
ionization energy \cite{lps,ab,pl}. This helps to explain the observed
\cite{ckesi} but previously unexplained (except by Eq.\ (\ref{vnequn})
below) closeness of these quantities.

Given $v(\re)$, the Hartree-Fock and exchange-only OEP orbitals and
other orbital energies differ in order $e^2$.  A good approximation to
the occupied Hartree-Fock orbitals can be achieved by a unitary
transformation of the occupied canonical OEP orbitals \cite{lhf}. In
particular, after mixing in the highest occupied OEP orbital, all the
canonical Hartree-Fock orbitals so approximated will decay
asymptotically with the same exponent. However, exact agreement is not
expected even to order $e^2$, since Eq.\ (\ref{psipert}) shows that to
order $e^2$ each {\it exact} canonical Hartree-Fock orbital is a
superposition of all the exact-exchange occupied {\it and unoccupied}
OEP orbitals. Thus, for a given $v(\re)$, the Hartree-Fock and
exchange-only OEP Slater determinants (which are invariant under
unitary transformations of their occupied orbitals) also differ to
order $e^2$, and only the Hartree-Fock determinant agrees to order
$e^2$ with the singles-only configuration interaction expansion of the
correlated many-electron wavefunction (Brillouin's theorem
\cite{lm2}).

\section{Iterative solution of the OEP equation}
\label{secconstruction}

\subsection{Constructing the OEP from the orbital shifts}

The fact that the OEP can be expressed solely in terms of the occupied
Kohn-Sham orbitals and their first-order shifts
%has not only formal but also far reaching practical consequences. 
leads to a simple scheme for its construction.
As a straightforward result from first-order perturbation theory, the partial
differential equations
%\begin{equation}
\begin{eqnarray}
\label{psipde}
\lefteqn{
(\hat{h}_{\mathrm{KS}\sigma}-\varepsilon_{i \sigma})\psi_{i \sigma}^*(\re)=
} \nonumber \\ & &
-[v_{\mathrm{xc}\sigma}(\re)-u_{\mathrm{xc}i\sigma}(\re)-
   (\bar{v}_{\mathrm{xc}i\sigma}-\bar{u}_{\mathrm{xc}i\sigma})
 ]\varphi_{i\sigma}^*(\re)
\end{eqnarray}
%\end{equation}
for the orbital shifts $\psi_{i \sigma}(\re)$ are obtained, where 
$
\bar{v}_{\mathrm{xc}i\sigma}=\int \varphi_{i\sigma}^*(\re) 
  v_{\mathrm{xc}\sigma}(\re)\varphi_{i\sigma}(\re) \, \mathrm{d}^3r
$
and
$
\bar{u}_{\mathrm{xc}i\sigma}=\int \varphi_{i\sigma}^*(\re) 
 u_{\mathrm{xc}i\sigma}(\re) \varphi_{i\sigma}(\re) \, \mathrm{d}^3r.
$
(Note that the same equations can be obtained by other arguments
\cite{kli2,graborev}). By solving Eq.\ (\ref{psipde}) for
$v_{\mathrm{KS}\sigma} \psi_{i\sigma}^*$, inserting the result into Eq.\
(\ref{oep}) multiplied by $v_{\mathrm{KS}\sigma}(\re)$, and solving
for $v_{\mathrm{xc}\sigma}$, the exact 
exchange-correlation potential can be expressed explicitly in terms of
only the occupied Kohn-Sham orbitals and their shifts
\cite{kli2,graborev}:
\begin{eqnarray}
\label{oepexpl1}
\lefteqn{
v_{\mathrm{xc}\sigma}(\re)=}
\nonumber \\ & &
\frac{1}{2 n_\sigma(\re)}
\sum_{i=1}^{N_\sigma}\left\{ 
\left| \varphi_{i\sigma}(\re) \right|^2 \left[ 
  u_{\mathrm{xc}i\sigma}(\re)+ 
  (\bar{v}_{\mathrm{xc}i\sigma}-\bar{u}_{\mathrm{xc}i\sigma}) \right]
\right.  \nonumber \\ & & \left. 
+\varphi_{i\sigma}(\re)\left[\left(
\frac{\hbar^2}{2m}\nabla^2+\varepsilon_{i\sigma}
\right)\psi_{i \sigma}^*(\re)\right] 
\right\} + c.c.
\end{eqnarray}
Invoking Eq.\ (\ref{oep}) once more, the last term can be rewritten to
obtain the expression from which the KLI approximation -- neglecting
the terms involving the $\psi_{i\sigma}^*(\re)$ -- has been justified
as a mean-field approximation \cite{kli2,graborev}:
\begin{eqnarray}
\label{oepexpl2}
\lefteqn{
v_{\mathrm{xc}\sigma}(\re)=}
\nonumber \\ & &
\frac{1}{2 n_\sigma(\re)}
\sum_{i=1}^{N_\sigma}\left\{ 
\left| \varphi_{i\sigma}(\re) \right|^2 \left[ 
  u_{\mathrm{xc}i\sigma}(\re)+ 
  (\bar{v}_{\mathrm{xc}i\sigma}-\bar{u}_{\mathrm{xc}i\sigma}) \right]
\right.  \nonumber \\ & & \left. 
-\frac{\hbar^2}{m}\nabla \cdot \left[\psi_{i \sigma}^*(\re)\nabla 
        \varphi_{i\sigma}(\re) \right] \right\} + c.c.
\end{eqnarray}
Since the potential $v_{\mathrm{xc}\sigma}(\re)$ of Eq.\ (\ref{oepexpl2}) is
fixed only up to an additive constant, one is 
at liberty to choose one of the constants
$\bar{v}_{\mathrm{xc}i\sigma}-\bar{u}_{\mathrm{xc}i\sigma}$ freely. 
The usual choice is \cite{kli2,graborev}
\begin{equation}
\label{vnequn}
\bar{v}_{\mathrm{xc}N_\sigma\sigma}-\bar{u}_{\mathrm{xc}N_\sigma\sigma}=0,
\end{equation}
to make the potential vanish at infinity. We will discuss this
point in greater detail in Section \ref{secasymptotics}.

Together with the orthogonality condition 
$\int \psi_{i\sigma}^*(\re) \varphi_{i\sigma}(\re) \, \mathrm{d}^3r=0$ 
that follows from Eq.\ (\ref{psipert}), Eq.\ (\ref{psipde}) uniquely
determines the orbital shifts. In case of degeneracies, the
orthogonality condition reads
$\int \psi_{i\sigma}^*(\re) \varphi_{j\sigma}(\re) \, \mathrm{d}^3r=0$
for all pairs $(i,j)$ with $\varepsilon_{i\sigma}=\varepsilon_{j\sigma}$.
In Ref.\ \onlinecite{oep1} it was demonstrated that the
$\psi^*_{i\sigma}(\re)$ can easily be calculated from Eq.\ (\ref{psipde}) and
used to construct the OEP. The basic idea is to combine the Kohn-Sham
equations (\ref{kseq}) with the orbital shift equations (\ref{psipde})
in a self-consistent iteration. This can be done in different ways.
The first and obvious one consists of the following steps: Solve the 
Kohn-Sham equations with an approximation to the OEP, e.g., do a KLI
calculation. Use the resulting $\varphi_{i\sigma}$ and
$v_{\mathrm{xc}\sigma}$ to construct the right-hand side of Eq.\
(\ref{psipde}). Solve, e.g., by conjugate gradient iteration
\cite{oep1,numrec}, for the orbital shifts. Insert the
$\psi_{i\sigma}^*$ into Eq.\ (\ref{oepexpl2}) to obtain an improved
approximation for $v_{\mathrm{xc}\sigma}$. Then self-consistently
solve the Kohn-Sham equations again for {\it fixed} $v_{\mathrm{xc}\sigma}$
to obtain a new total energy and new eigenvalues. Repeat these steps
until convergence is achieved.
 
This simple procedure only involves the occupied orbitals and their
shifts. Instead of an integral equation, it only requires solving
partial differential equations, which can easily be done. We thus gain
all the advantages of avoiding the explicit evaluation of the response
function that are known from other areas of linear response theory
\cite{sternheimer,baroni}.

In practice, a small modification of this algorithm is useful. After
$v_{\mathrm{xc}\sigma}$ has been updated, one need not directly go
back to the Kohn-Sham equations. Instead, one can solve Eq.\
(\ref{psipde}) again, keeping the left-hand side and the
$\varphi_{i\sigma}$ fixed but evaluating the right-hand side with the
new $v_{\mathrm{xc}\sigma}$. Using this additional
($\psi_{i\sigma}^*,v_{\mathrm{xc}\sigma}$)-cycle speeds up the
convergence, and Eq.\ (\ref{psipde}) is easier to solve than Eq.\
(\ref{kseq}) since it is just a differential and not an eigenvalue
equation. We discuss numerical aspects of the iteration and its
convergence in detail in Appendix \ref{appendixc}. Here we just point
out that the iteration converges very quickly: Starting , e.g. for the
Be atom, from the KLI approximation and using 5 cycles per iteration,
the total energy is correct after the first iteration and all
eigenvalues after 5 iterations.

The iterative approach is also accurate. One criterion that
indicates an accurate OEP can be inferred directly from Eq.\
(\ref{oep}). The function
\begin{equation}
\label{defs}
S_\sigma(\re)=\sum_{i=1}^{N_\sigma}
\psi_{i\sigma}^*(\re)\varphi_{i\sigma}(\re)+c.c.
\end{equation}
must vanish for the exact OEP. Therefore, the smaller the maximum
value $S_\mathrm{max}$ that $S(\re)$ takes in a given region of
space, the more accurate is the potential. For the KLI potential of
the Be atom, we find $S_\mathrm{max}=0.03 \, a_0^{-3}$. After 5
iterations with 5 cycles, $S_\mathrm{max}$ has fallen below $10^{-6}
\, a_0^{-3}$. A second accuracy criterion is the exchange virial
relation \cite{virial,wpcms}. It is slightly violated by the KLI
approximation with a relative error of about 1\%. By the time our
iteration has refined all eigenvalues to 0.0001 hartree (0.1 mhar)
accuracy, the error in the virial relation has been reduced to
$10^{-4}$\%, and further iteration reduces it to $10^{-7}$\%. The
method is thus at least as accurate as the direct solution of the
integral equation \cite{engel1,engel2}.

However, applied to finite systems, Eq.\ (\ref{oepexpl2}) has an
unpleasant feature: The second 
term under the sum on the right-hand side must be divided by the
density. While this is not a problem theoretically, it complicates the
numerical evaluation of Eq.\ (\ref{oepexpl2}) for finite systems. Far
away from the system the density decays exponentially, and numerically
dividing by a rapidly vanishing function introduces
inaccuracies. These inaccuracies in the asymptotic region can magnify
during the iteration and, if not taken care off, can spoil the whole
scheme. In Appendix \ref{appendixc} we discuss this problem in greater
detail, together with strategies to overcome it. Here we want
to focus on how it can be completely avoided.

For many numerical problems, iterative solution strategies
can be designed in different ways \cite{numrec}. Therefore, there is
no reason to believe that Eq.\ 
(\ref{oepexpl2}) is the only way that the information contained in the
orbital shifts can be used. In fact, Eq.\ (\ref{defs}) suggests a very
simple, pragmatic alternative: Since $S_\sigma(\re)$ is an
indicator for the error inherent in a given approximation to the OEP,
we improve the given approximation by adding the corresponding
$S_\sigma(\re)$ to it:
\begin{equation}
\label{siter}
v_{\mathrm{xc}\sigma}^\mathrm{new}(\re)=
v_{\mathrm{xc}\sigma}^\mathrm{old}(\re)+c S_\sigma(\re).
\end{equation}
The real parameter $c>0$ with dimension energy times volume is
introduced because $S_\sigma(\re)$ is not an 
exact representation of the error but just an estimate. 
Since it is not 
guaranteed that the $v_{\mathrm{xc}\sigma}^\mathrm{new}(\re)$ from
Eq.\ (\ref{siter}) will obey Eq.\ (\ref{vnequn}), we enforce it
explicitly by adding the constant
$\bar{u}_{\mathrm{xc}N_\sigma\sigma}-\int \varphi_{N_\sigma}^*(\re)
v_{\mathrm{xc}\sigma}^\mathrm{new}(\re)\varphi_{N_\sigma\sigma}(\re)
\, \mathrm{d}^3r $ to $v_{\mathrm{xc}\sigma}^\mathrm{new}$ each time
it has been updated via Eq.\ (\ref{siter}). (See Ref.\
\onlinecite{oep1} for an intuitive explanation of the same idea.)

Eq.\ (\ref{siter}), together with adding the constant, can replace the
update of $v_{\mathrm{xc}\sigma}$ via Eq.\ (\ref{oepexpl2}), and all
the other steps described in the three paragraphs following Eq.\
(\ref{vnequn}) stay the same.
As shown below, introducing $c$ can be understood as replacing by a constant a
term that requires dividing by the density. Doing so, we drop
some information from the error feedback, and therefore the iteration
converges somewhat slower than the direct iteration of Eq.\
(\ref{oepexpl2}). But with 10
($\psi_{i\sigma}^*,v_{\mathrm{xc}\sigma}$)-cycles per iteration, the
total energies of the atoms Be, Ne, Na, Mg, and Ar converge within 0.1
mhar in 2, 2, 2, 5, and 7 iterations, respectively (and a larger
number of cycles further reduces the number of necessary
iterations). The computational effort is therefore only moderate. The
OEP energies from our solution are identical to the ones found from
the considerably more involved direct solution of the integral
equation \cite{talman,graborev}, and the iteration is as accurate (see
Appendix \ref{appendixc} for tests).

This approach can be motivated further
by rewriting Eq.\ (\ref{psipde}) such that the left-hand
side becomes 
$(\hbar^2/(2m) \nabla^2 +\varepsilon_{i\sigma})\psi_{i\sigma}^*$ and
inserting the result into the right-hand side of Eq.\
(\ref{oepexpl1}) to obtain
\begin{equation}
\label{motivate}
v_{\mathrm{xc}\sigma}(\re)=v_{\mathrm{xc}\sigma}(\re)+
\frac{v_{\mathrm{KS}\sigma}(\re)}{2 n_\sigma(\re)}S_\sigma(\re).
\end{equation}
This is a trivial identity if $\varphi_{i\sigma}$ and
$\psi_{i\sigma}^*$ are the orbitals and shifts corresponding to the
true OEP. However, if $\varphi_{i\sigma}$ and $\psi_{i\sigma}^*$ have
been calculated from an approximation to the OEP, then the second term
on the right-hand side does not vanish and represents
an error term. Its most important part
is the function $S_\sigma(\re)$ which by definition
contains the information about where the OEP equation is violated. By
subtracting the error term from the approximation, we get a better
approximation. But we only want to subtract the part
that is proportional to $v_{\mathrm{xc}\sigma}$,
\begin{equation}
\label{nearlythere}
v_{\mathrm{xc}\sigma}^\mathrm{new}(\re)=
v_{\mathrm{xc}\sigma}^\mathrm{old}(\re)
-\frac{v_{\mathrm{xc}\sigma}^\mathrm{old}(\re)}{2 n_\sigma(\re)} 
S_\sigma(\re),
\end{equation}
because for stability reasons we do not want to change
$v_{\mathrm{xc}\sigma}$ too drastically in each
iteration. (Subtracting the full term using $v_\mathrm{KS}$ would,
e.g., introduce a very different length scale and the nuclear
singularity into the iteration. Besides, in the neutral systems of
interest here, $v_\mathrm{xc}$ dominates the asymptotics of
$v_\mathrm{KS}$.)  We have tested Eq.\ (\ref{nearlythere}) for
spherical atoms, and it converges nearly as quickly as the direct
iteration of Eq.\ (\ref{oepexpl2}). But since Eq.\ (\ref{nearlythere})
also requires dividing by the density, the simpler iteration based on
Eq.\ (\ref{siter}) is more practical, in particular for the cluster
calculations.  Eq.\ (\ref{nearlythere}) provides an interpretation for
the constant $c$: It takes the role of an average value for
$v_{\mathrm{xc}\sigma}^\mathrm{old}(\re)/(2n_\sigma(\re))$. Since
$v_{\mathrm{xc}\sigma}$ is negative and we chose $c$ to be positive,
the signs in Eqs.\ (\ref{siter}) and (\ref{nearlythere}) are opposite.
In practice, the numerical value of $c$ can be determined
unproblematically by trial and error. For the systems we studied, we
found $c$ close to the values inferred from Eq.\ (\ref{nearlythere});
see Appendix \ref{appendixc}.

\subsection{OEP for Na clusters}

Eq.\ (\ref{siter}) can very easily be employed in three-dimensional
calculations and allows us to study the influence that the exact exchange
functional has on non-spherical systems. The electrons
in atoms are rather strongly localized. Metal clusters, and in
particular sodium clusters, have a very different electronic structure
with strongly delocalized valence electrons. They can thus be regarded
as an ``opposite'' test case for the influence of exact exchange. Of
particular interest is their static electric polarizability, which we
will address in Section \ref{secpol} below.
\begin{table*}[hbt]
\caption{Absolute value of the total energy and lowest and highest
occupied Kohn-Sham eigenvalue for small sodium clusters, in hartree, for
exchange only OEP, KLI, and local exchange, and for local exchange and
correlation \cite{pw}. For the two-valence-electron system
$\mathrm{Na}_2$, the KLI potential is the exact OEP.}   
\label{tab1}
\begin{ruledtabular}
\begin{tabular}{c|cc|ccc|ccc|ccc|ccc}
& \multicolumn{2}{c}{$\mathrm{Na}_2$ } & \multicolumn{3}{c}{$\mathrm{Na}_4$} &
  \multicolumn{3}{c}{$\mathrm{Na}_6$} & \multicolumn{3}{c}{$\mathrm{Na}_8$} &
  \multicolumn{3}{c}{$\mathrm{Na}_{10}$} \\ \hline
& -E & $-\varepsilon_L$ = $-\varepsilon_H$ 
& -E & $-\varepsilon_L$ & $-\varepsilon_H$
& -E & $-\varepsilon_L$ & $-\varepsilon_H$
& -E & $-\varepsilon_L$ & $-\varepsilon_H$
& -E & $-\varepsilon_L$ & $-\varepsilon_H$ \\ \hline
xOEP  & 0.3768 & 0.1701 
      & 0.7531 & 0.1762 & 0.1401 
      & 1.1421 & 0.1861 & 0.1496
      & 1.5285 & 0.1954 & 0.1459
      & 1.9082 & 0.1956 & 0.1263 \\ \hline
xKLI  &  - & - 
      & 0.7528 & 0.1776 & 0.1394 
      & 1.1417 & 0.1871 & 0.1494
      & 1.5282 & 0.1963 & 0.1458
      & 1.9070 & 0.1965 & 0.1248   \\ \hline
xLDA  & 0.3479 & 0.0874 
      & 0.7047 & 0.1123 & 0.0717 
      & 1.0810 & 0.1205 & 0.0824
      & 1.4619 & 0.1331 & 0.0826
      & 1.8257 & 0.1410 & 0.0664  \\ \hline
LDA   & 0.4018 & 0.1135 
      & 0.8166 & 0.1384 & 0.0976    
      & 1.2511 & 0.1469 & 0.1085
      & 1.6926 & 0.1596 & 0.1085              
      & 2.1169 & 0.1672 & 0.0926                   
\end{tabular}
\end{ruledtabular}
\end{table*}

\begin{figure}[h]
\includegraphics[width=8.1cm]{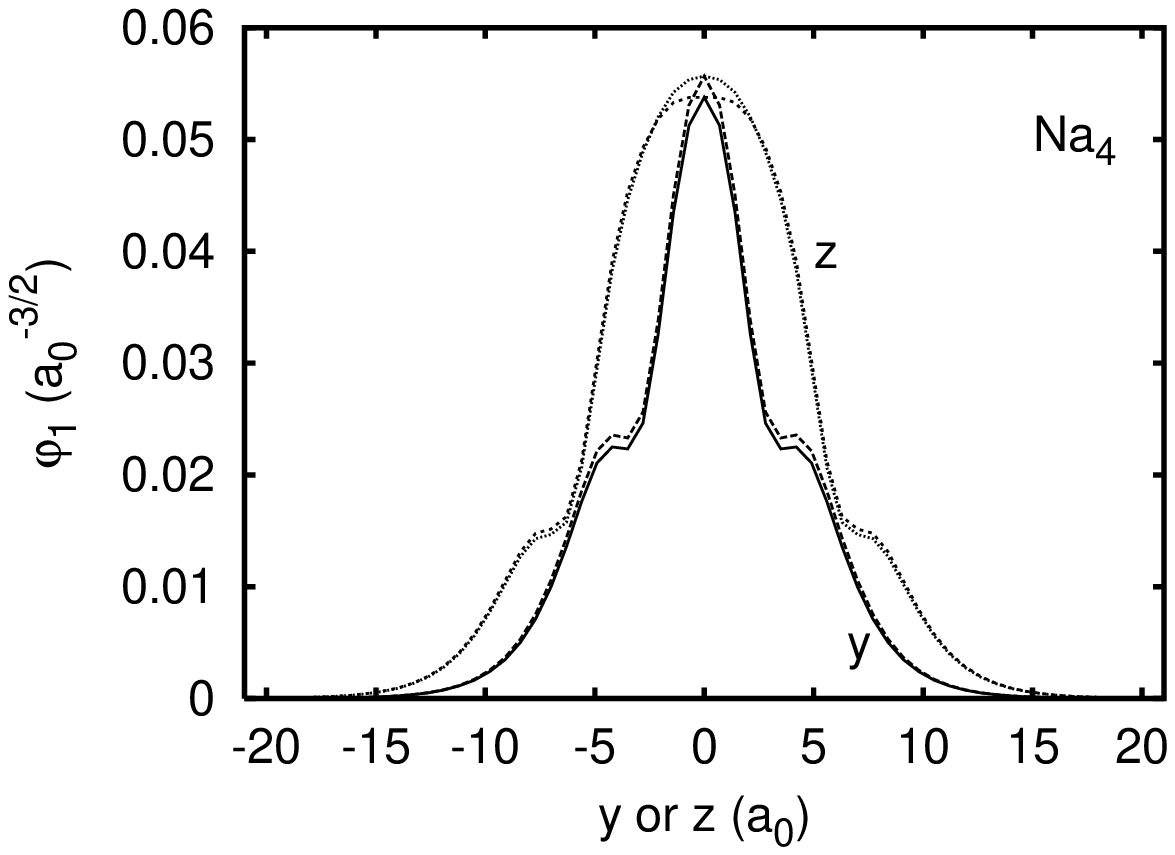}
\includegraphics[width=8.1cm]{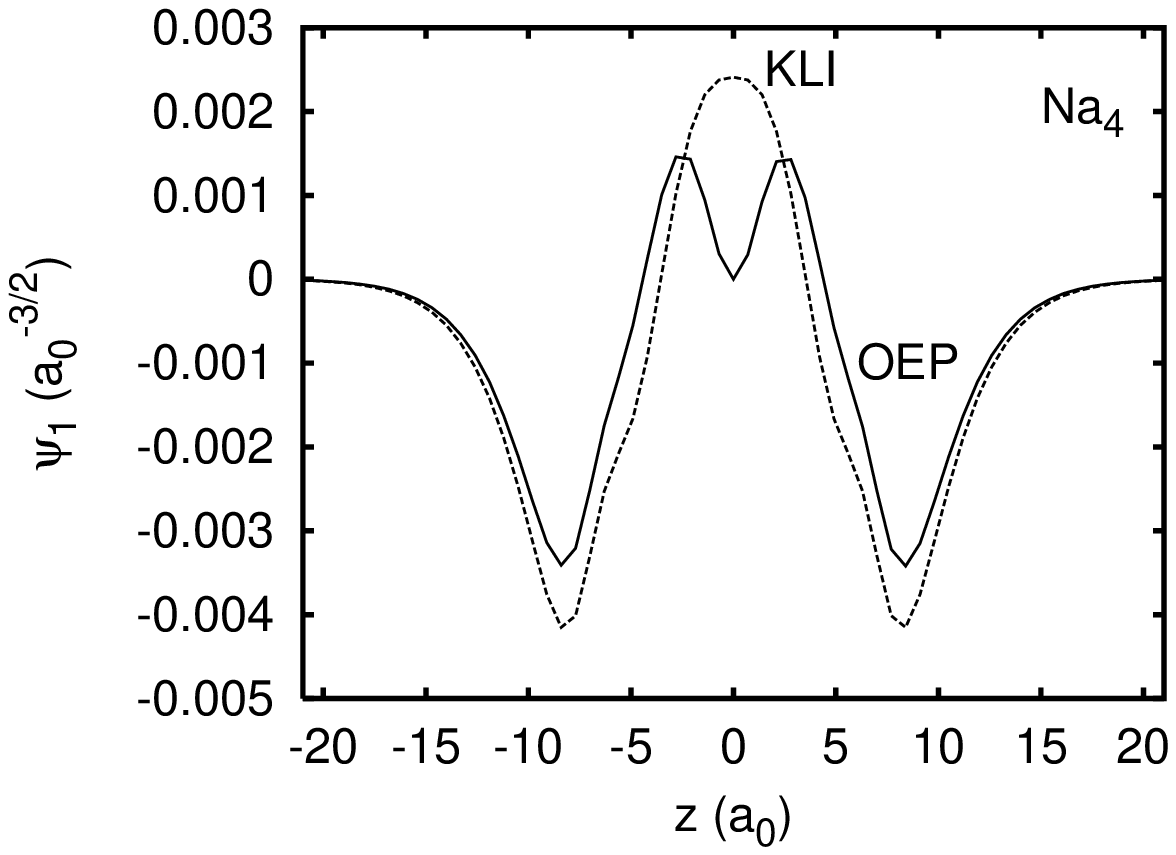}\\
\includegraphics[width=8.1cm]{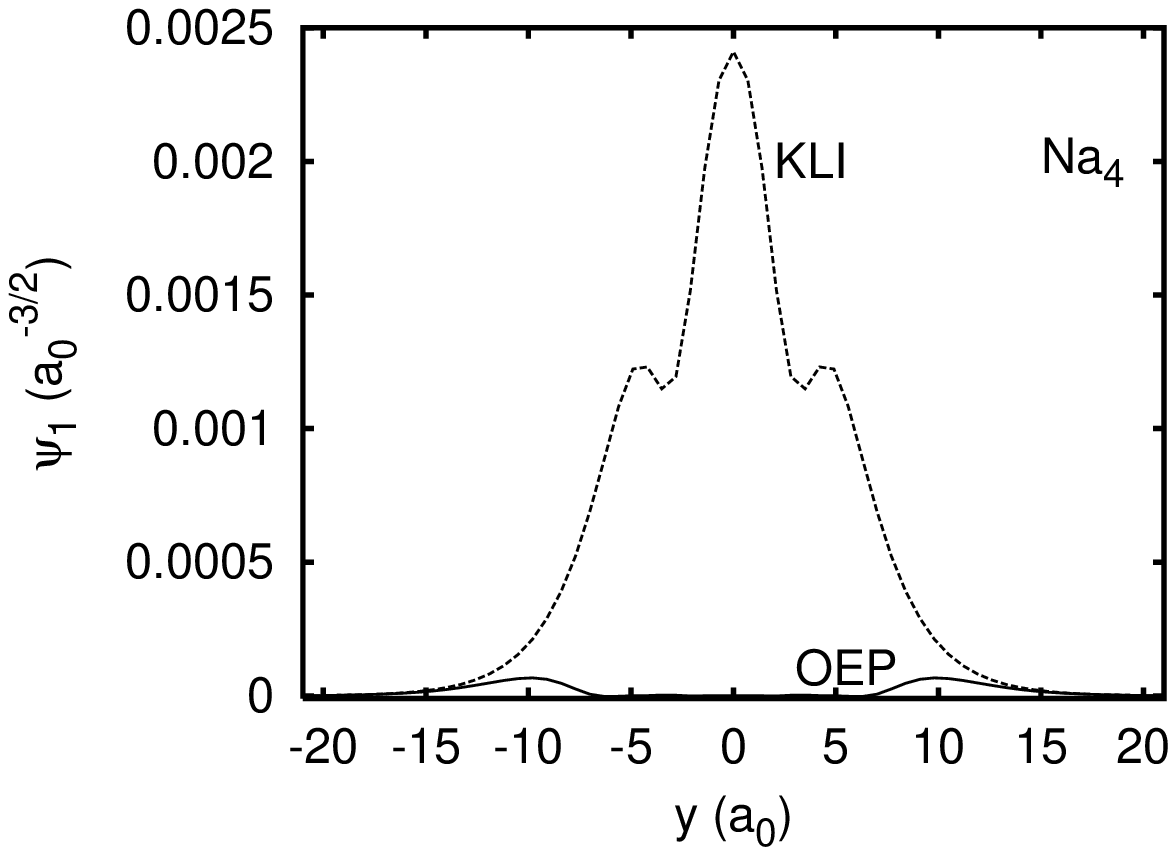}\\
\caption{The lowest Kohn-Sham orbital and corresponding orbital shift for
$\mathrm{Na}_4$. Top: Full and long dashed: Orbital from OEP and
KLI potential, respectively, along the y-axis. Short dashed and dotted:
Orbital from OEP and KLI potential, respectively, along the z-axis.
Middle: The corresponding orbital shifts plotted along the
z-axis. Full: OEP, dashed: KLI. Bottom: Same orbital shifts
along the y-axis. Full: OEP, dashed: KLI. See text for discussion.
\label{fig1}} 
\end{figure}
We first calculated the ground-state properties of small sodium
clusters using the exact exchange functional together with the exact
OEP and with the KLI approximation, and for comparison also with
LDA for exchange only (xLDA) and LDA for exchange and correlation
\cite{pw}. For reasons discussed in Section \ref{secpol}, we
used the optimized cluster geometries and pseudopotential from Ref.\
\onlinecite{epjd,prb}. The results are summarized in Table
\ref{tab1}. For the
smallest cluster $\mathrm{Na}_2$, which has one valence electron of
each spin, the KLI approximation is exact. This is confirmed by the
exchange virial relation: Whereas for all other clusters the virial
relation is slightly violated by the KLI approximation with a relative
error of about 0.5\%, its relative error is only $10^{-6}$ for
$\mathrm{Na}_2$. The total energies of all larger clusters are lower
with the OEP than with the KLI potential, as required. The absolute
average difference of 0.5 mhar indicates that KLI is a rather good
approximation for the total energies. It should also be noted,
however, that the relative error is considerably larger for the
clusters than for the atoms. The differences between the Kohn-Sham
eigenvalues obtained from OEP and KLI are larger than the differences
in total energy, and there is a clear pattern: As for the atoms, the
OEP raises the lowest occupied eigenvalues and lowers the highest
occupied ones. Thus,  
the energy range of the occupied Kohn-Sham spectrum is up to 6\%
narrower in OEP than in KLI. This might well influence time-dependent
DFT calculations to which the eigenvalues are an important
ingredient. Comparing the results from the exact exchange functional
to the xLDA results shows that, as expected, exact and local exchange
lead to considerably different eigenvalue spectra: Exact exchange
eigenvalues are more negative, and in particular the highest one is
relatively much more negative. The total energy in xLDA is less
negative than that with exact exchange, reflecting the well known
effect that the local approximation underestimates the magnitude of
the exchange energy. When local correlation is included, the local
exchange error in the total energy is compensated to a good part. The
eigenvalues, however, are still considerably smaller in absolute value
even with correlation.

Finally, we investigate the orbital shifts
$\psi_{i\sigma}^*(\re)$. This can be done in a transparent way for
$\mathrm{Na}_4$: With two electrons of each spin, the KLI potential
and OEP are already non-trivially different, yet we only need to look
at two orbitals and shifts. The upper part of Fig.\ \ref{fig1} shows
the lower of the two Kohn-Sham orbitals as obtained from the KLI
approximation and the OEP. The orbital is plotted for both potentials
once along the y-axis and once along the z-axis. The ionic
configuration of $\mathrm{Na}_4$ and our labeling of axes is shown in
Fig.\ \ref{fig2}. Clearly, the orbital obtained within KLI is very
similar to the orbital from the exact OEP. The second Kohn-Sham
orbital which is not shown in Fig.\ \ref{fig1} can be described as a
$p_z$ orbital, i.e., it has a node in the x-y plane. The second
KLI and OEP orbitals are also very similar. However, the conclusion that,
therefore, the orbital shifts obtained from KLI and OEP should be very
similar, too, is wrong. In the middle part of Fig.\ \ref{fig1} the
first orbital shift is shown along the z-axis. The shift calculated
from the OEP via Eq.\ (\ref{psipde}) shows a ``dip'' at the origin
which is not seen in the shift corresponding to the KLI potential. And
along the y-axis (lower part of Fig.\ \ref{fig1}), the shifts obtained
from the KLI potential and the OEP are totally different: The KLI
shift shows a pronounced central peak, whereas the OEP shift
practically vanishes on the y-axis. The explanation for these somewhat
surprising observations can be found directly in the OEP equation Eq.\
(\ref{oep}). As mentioned before, the x-y plane is a nodal plane for
the higher of the two Kohn-Sham orbitals of $\mathrm{Na}_4$. The lower
orbital, however, does not vanish in the x-y-plane. Eq.\ (\ref{oep})
can therefore only be fulfilled in the x-y-plane if the orbital shift
corresponding to the lower orbital vanishes in this plane. The KLI
potential is only an approximation to the OEP, and therefore need not
and, as seen in Fig.\ \ref{fig1}, does not fulfill Eq.\ (\ref{oep}) in
the x-y-plane. The iteratively constructed OEP, in contrast, makes the
orbital shift vanish in the plane, as it should. The lower part of
Fig.\ \ref{fig1} thus once more confirms that we are indeed
constructing the exact OEP. The reason for the pronounced difference
between the OEP and KLI orbital shift is discussed in greater detail
in Appendix \ref{appendixd}.   

It remains to be explained why the orbital shift shown in the lower
part of Fig.\ \ref{fig1} vanishes perfectly between y=-7 $a_0$ and
z=+7 $a_0$, but still shows a tiny bump at larger distances, a remnant
from the KLI orbital shift. The reason is our iteration based on Eq.\
(\ref{siter}). It corrects the potential quickly in the
energetically-important region where the orbitals and their shifts are
appreciable. In the asymptotic region, however,
$\varphi_{i\sigma}(\re)$ and $\psi_{i\sigma}^*(\re)$ decay
exponentially, and consequently the potential is modified to a lesser
extent. But why does this matter since it is believed that the KLI
potential is asymptotically correct? The answer is that the KLI
potential is asymptotically correct everywhere {\it except} on nodal
surfaces of the highest-occupied orbital that extend out to
infinity. On such nodal surfaces, a small difference between KLI
potential and OEP is observed even in the asymptotic region, and it is
this difference which gives rise to the above mentioned ``bump''. This
aspect will be discussed in greater detail in the next section. Here
we just briefly want to mention that, if needed, the iteration based
on Eq.\ (\ref{siter}) can be modified to yield quickly the correct
asymptotic behavior also on the nodal surface: the constants
$\bar{v}_{\mathrm{xc}i\sigma}-\bar{u}_{\mathrm{xc}i\sigma}$ which
determine the asymptotic behavior (see below) only depend on the
energetically-important region. Their values (and thereby the
asymptotic limits of the exchange potential) are thus accurately known
after just a few iterations, even when the potential converges slowly
in the asymptotic region.

\section{Asymptotic behavior of $v_\mathrm{x}(\re)$}
\label{secasymptotics}

The long-range asymptotic behavior of the exact exchange OEP can be
inferred from Eq.\ (\ref{oepexpl2}). The first term in the sum over
all occupied orbitals determines the asymptotics
\cite{graborev,goerasymp}, since the highest-occupied orbital shift
falls off faster than the highest-occupied orbital. The Kohn-Sham
orbitals of a finite system decay exponentially like $\exp[-(- 2
m\varepsilon_{i\sigma}/\hbar^2 )^{1/2}r]$, where $r$ is the distance
from the system's center. This is a consequence of the locality of the
Kohn-Sham potential and means that each orbital's decay is dominated
by its own orbital energy. Therefore, the sum asymptotically will be
dominated by the slowest-decaying orbital, which is the one with the
highest orbital energy. Since $u_{\mathrm{x} N_\sigma \sigma}
\stackrel{r\rightarrow\infty}{\longrightarrow}-e^2/r$,
$v_\mathrm{x}(\re)$ for a finite system goes to zero like $-e^2/r$ in
all regions of space that are dominated by the highest-occupied
orbital density $n_{N_\sigma}(\re)=|\varphi_{N_\sigma}(\re)|^2$. That
the potential goes to zero and not some finite constant is due to Eq.\
(\ref{vnequn}).
%, and we can choose this condition because any potential is fixed
%only up to one additive constant.  
However, there may also be asymptotic regions of space where
$n_{N_\sigma}(\re)$ vanishes, but the orbital density of a lower
occupied orbital $M_\sigma$ does not. As pointed out in Ref.\
\onlinecite{goerasymp}, this situation occurs whenever the highest
occupied orbital has a nodal surface that extends out to infinity. In
these regions, one can go through all the arguments used to derive the
asymptotic behavior and replace $n_{N_\sigma}(\re)$ by
$n_{M_\sigma}(\re)$. But the reference point for the potential was
already fixed via Eq.\ (\ref{vnequn}). We cannot choose a second one,
therefore there is no equivalent of Eq.\ (\ref{vnequn}) for the
$M_\sigma$ orbital. Consequently, the potential will asymptotically go
to
$C=\bar{v}_{\mathrm{xc}M_\sigma\sigma}-\bar{u}_{\mathrm{xc}M_\sigma\sigma}$,
where in general $C \neq 0$. Thus, it must be concluded that the exact
Kohn-Sham exchange potential can go to different asymptotic constants
in different spatial directions.

In Ref.\ \onlinecite{kli2} this aspect of Eq.\ (\ref{oepexpl2}) was
first discussed, but its consequences for the Kohn-Sham potentials of
real systems were not considered. In Ref.\ \onlinecite{goerasymp} it was
demonstrated that the effect indeed occurs for real systems, but
the numerical calculations were restricted to the "localized Hartree-Fock"
approximation to the OEP. Finally, in Ref.\ \onlinecite{oep1} and in this
paper, we evaluate the exact OEP including the asymptotic region
for realistic, three-dimensional systems and confirm the surprising
asymptotic behavior of the exact exchange potential.

\begin{figure}[t]
\includegraphics[width=8.1cm]{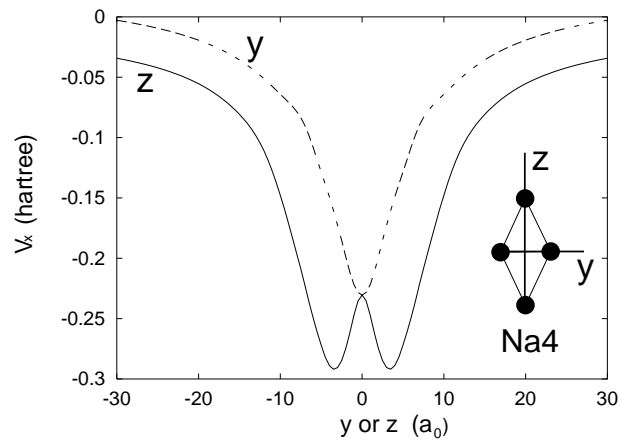}
\caption{The exact exchange potential of $\mathrm{Na}_4$ in hartree
along the z-axis (full line) and the y-axis (dashed line) in bohr
$a_0$. The potential goes to different asymptotic limits in different
directions.
\label{fig2}} 
\end{figure}
Fig.\ \ref{fig2} shows the exchange-only OEP for the cluster
$\mathrm{Na}_4$, the electronic structure of which was discussed in
the previous section. The full line shows $v_\mathrm{x}(\re)$ along
the z-axis. In this direction, the density is asymptotically dominated
by the highest occupied orbital and the potential falls off like
$-e^2/r$, as expected. The dashed line shows $v_\mathrm{x}(\re)$ along
the y-axis. Clearly, the potential falls of in the same way, but tends
to a different constant $C$. We find $C=0.0307$ har. This
non-vanishing value for $C$ is of course found only in the
x-y-plane. Moving away from the origin at any low angle to this nodal
plane, one eventually reaches a region where the highest-occupied
orbital will dominate the density and $v_\mathrm{x}(\re)$ will tend to
zero. A lower angle just means that the asymptotic region starts
further away from the center of the cluster. But nevertheless, since
the exchange potential is a continuous function of $\re$, the
non-vanishing asymptotic constant has a substantial influence on
$v_\mathrm{x}(\re)$ even outside of the nodal plane of the highest
occupied orbital.
\begin{figure}[b]
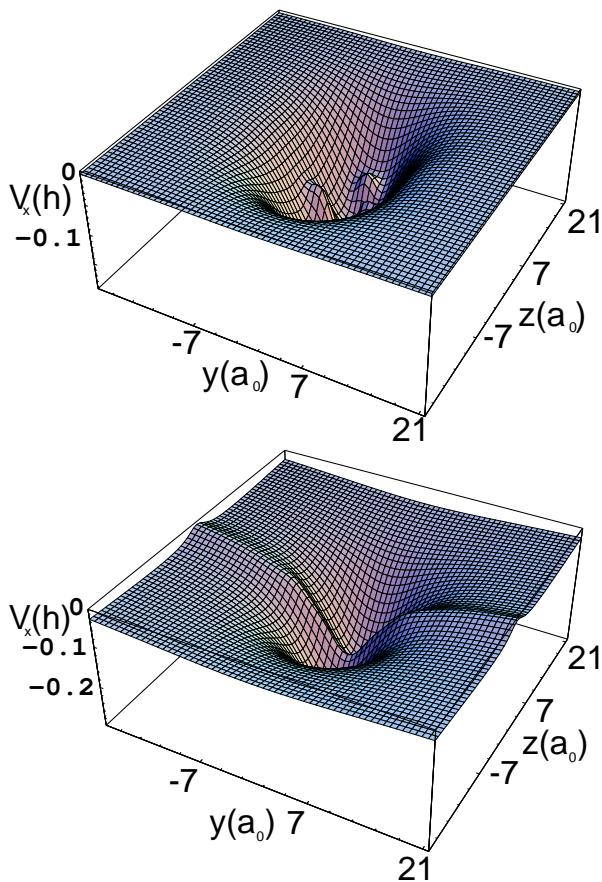

\includegraphics[width=8.1cm]{Na4potxLDA.epsi}\\
\includegraphics[width=8.1cm]{Na4potxOEP.epsi}
\caption{The exchange potential ``landscape'' for $\mathrm{Na}_4$ in
the y-z plane: $v_\mathrm{x}(\re)$ in hartree defines the
``height'' at each spatial point (y,z) for fixed
x=0. Upper: exchange-only LDA potential; lower: exact exchange
potential (OEP). Note 
%the clearly visible locality of the LDA potential and 
the greater depth, slower decay, and
pronounced ``ridge'' seen in the OEP. 
\label{fig3}} 
\end{figure}
To demonstrate this, we show in the lower half of Fig.\ \ref{fig3} the
exact exchange potential for $\mathrm{Na}_4$, and the potential from
the local density approximation in the upper. For each point in the
y-z-plane, the potential we calculated on a numerical grid is plotted
as the ``height''. The LDA-potential mirrors the density closely, so
the ionic cores which cause ``bumps'' in the valence electron density
show up in the LDA potential. We also observe the well-known effect
that the LDA potential decays very quickly: it vanishes on the
boundary of the plotted region, whereas the exact exchange potential
still takes a non-vanishing value there. But the most obvious
difference between the exact and the local exchange potential is that
the former shows a pronounced ``ridge'' running along the y-axis. This
ridge is a consequence of the non-vanishing asymptotic constant: The
potential has to go up to reach its correct value on the nodal
surface. We find qualitatively the same behavior for all the clusters
we studied here. For $\mathrm{Na}_6$ and $\mathrm{Na}_8$, there are
two degenerate highest orbitals, and the OEP goes to $C=0.0378$ and
$C=0.0084$, respectively, on the z-axis. (We choose our Cartesian
coordinates such that the axes are the principal axes of the cluster's
tensor of inertia, and z is the axis whose principal value differs
most from the average. All constants are in har.) The relatively small
constant for the ``magic'' $\mathrm{Na}_8$ reflects the fact that the
highest orbital is nearly degenerate with the two lower-lying
ones. For $\mathrm{Na}_{10}$, the situation is similar to
$\mathrm{Na}_4$, with $C=0.0107$ in the x-y-plane. Finally, it is
clear from Eq.\ (\ref{oepexpl2}) that the asymptotic constants are
qualitatively correct in the KLI potential. But KLI and OEP orbitals
differ slightly, as shown in the previous section. Consequently, we
find constants that differ by about $0.001$ har in all cases we
investigated.

The non-vanishing asymptotic constants are thus the rule, not the
exception, and one may ask how this result can be true since there is
a physical argument to explain the asymptotic behavior of the
Kohn-Sham potential: An electron that wanders far out from a finite
neutral system basically sees the attraction from the hole it leaves
behind. Thus, the Kohn-Sham potential should fall of to $-e^2/r$
everywhere, without additional constants \cite{hs}. However, appealing as this
argument is, it cannot be applied to the Kohn-Sham potential. The
Kohn-Sham potential is not a physical potential whose value we can
measure. It is a mathematical construction, and therefore can
have properties that physical potentials do not show. But of course
the Kohn-Sham potential is a very clever mathematical construction,
designed to yield from a simple procedure an accurate description of
complicated quantities we can measure, like the ground-state energy of
a many body system. Therefore, it is important to understand its
features, in particular the surprising ones. The first and probably
most extreme is the discontinuity of the Kohn-Sham potential as a
function of electron number \cite{pplb,kli1,gb}. The existence of
non-vanishing asymptotic constants is a second one.

\section{Static electric polarizabilities from the exact exchange
functional}
\label{secpol}

To demonstrate the importance of incorporating exact exchange into
functionals, we investigate the static electric dipole polarizability.
There has been a long-standing controversy about the influence of the
long-range asymptotics of the Kohn-Sham potential and the
self-interaction error on the optical properties of metal clusters
\cite{epjd,moullet,pacheco,guetxc,guan,alonso,vasiliev,ullrich,marques,gisbergen}.
The static electric dipole polarizability has been of particular
interest: It was one of the earliest observables to give insight into
the structure of clusters \cite{knight}, its measurement gives
information on the effects that govern cluster growth \cite{rayane},
and it may be used as an indicator for whether the bulk limit is
reached \cite{kresin}. The polarizability of small sodium clusters has
been calculated before
\cite{epjd,moullet,guan,vasiliev,marques,gisbergen}. It was found that, 
while local and semi-local density functionals yield
polarizabilities in reasonable qualitative agreement with experiment,
they do not lead to quantitative agreement. However, it has also been
demonstrated \cite{prl,leeor,guetpol} that a finite cluster
temperature greatly influences the static electric polarizability,
mainly through the effect of thermal expansion \cite{prl}. The
clusters' temperature in the experiments is presently not accurately
known, but it is estimated to be high enough to account perhaps for
all of the observed difference \cite{epjd,prl,leeor}. Nevertheless, it
is of great practical interest to obtain accurate theoretical values
for the polarizability because the polarizability could then be used
as a ``thermometer'' to estimate experimental cluster temperatures, as
discussed in greater detail by Kronik and co-workers \cite{kronik}.

Our exact-exchange OEP calculations allow us to answer some of the
questions related to the influence of self-interaction errors and
long-range asymptotics that have been raised in the past. In the present
study, our focus is on investigating these two effects by comparing LDA and
exact-exchange results. In order to sort out the effects as clearly as
possible, we want the valence electrons to move in exactly the same external
potential for all $E_\mathrm{xc}$ functionals. Thus, we use the same
pseudopotential and cluster geometries for different $E_\mathrm{xc}$. For an
accurate comparison with experimental data, a consistent exact-exchange
pseudopotential \cite{bylander,oeppp1,oeppp2,oeppp3} should of course
be used and the cluster structures re-optimized.
% for the $E_\mathrm{x}^\mathrm{ex}$ calculations. 
But the comparison with experiment would also require inclusion of correlation
effects. Since a correlation functional generally compatible with full
exact exchange is not yet available, these aspects will have
to be addressed in future work.

We calculated the mean static electric dipole polarizability
$
\bar{\alpha}=\mathrm{tr}(\mbox{\boldmath$\alpha$})/3
$
for the neutral sodium clusters with even electron numbers between 2
and 10, using the LDA-optimized geometries of Refs.\ \onlinecite{epjd,prb}. 
The polarizability tensor \mbox{\boldmath$\alpha$} was obtained
from the electric dipole moment 
$
\label{defdipm}
\mu_j({\bf F})=-e \int r_j n({\bf r, F})\,d^3r 
%+ e Z \sum_{k=1}^N R_{k\,j}
$
by evaluating the definition 
$
%\begin{equation}
\label{defaij}
%\alpha_{ij}=\lim_{F_i \rightarrow 0}\frac{\partial \mu_j}{\partial F_i}
\alpha_{ij}=\lim_{F_i \rightarrow 0}\partial \mu_j / \partial F_i
%  \rightarrow
%  \frac{\mu_j(+F_i)-\mu_j(-F_i)}{2 F_i}, 
%\hspace{0.5cm}  
, \, \,
i,j=x,y,z,
%\end{equation}
$
for a small but finite electrical field $\mathbf F$ (see Ref.\
\onlinecite{epjd} for details, e.g, on the role of the ionic cores). 

\begin{table}[b]
\caption{Mean static electric dipole polarizability $\bar{\alpha}$ of
  Na clusters in $\AA^3$ 
  for different approximations to $E_\mathrm{xc}$ and $v_\mathrm{xc}$. Columns
  from left to right: exact exchange with exact OEP; exact exchange with
  potential in KLI approximation; LDA for exchange only; LDA \cite{pw} for
  exchange and correlation; relative difference in \% between local and
  exact-exchange polarizabilities,
$\Delta\alpha=\bar{\alpha}^{\mathrm{x,LDA}}-\bar{\alpha}^{\mathrm{x,OEP}}$.}
\label{tab2}
\begin{ruledtabular}
\begin{tabular}{cccccc}
Cluster& $E_\mathrm{x}^\mathrm{ex,OEP}$ & $E_\mathrm{x}^\mathrm{ex,KLI}$ &
$E_\mathrm{x}^\mathrm{LDA}$ & $E_\mathrm{xc}^\mathrm{LDA}$ &
$\Delta \alpha/\bar{\alpha}^{x\mathrm{OEP}}$ \\ \hline
$\mathrm{Na}_2$     &  36.9 &  36.9 &  40.4 &  37.1 & 9.5 \\
$\mathrm{Na}_4$     &  78.7 &  79.4 &  85.7 &  78.8 & 8.9 \\ 
$\mathrm{Na}_6$     & 107.6 & 108.3 & 115.0 & 107.3 & 6.7 \\
$\mathrm{Na}_8$     & 125.8 & 126.6 & 131.4 & 123.0 & 4.5 \\
$\mathrm{Na}_{10}$  & 158.7 & 160.8 & 168.4 & 157.3 & 6.1
\end{tabular}
\end{ruledtabular}
\end{table}
Table \ref{tab2} summarizes the results of our calculations. We find
that, for exact exchange and no correlation, the polarizabilities
calculated using the exact OEP are consistently lower than the ones
calculated from the KLI approximation. This is in agreement with our
expectations, since the OEP yields lower total energies, i.e, binds
the valence electrons more strongly than the KLI potential, and the valence
electron density is thus harder to displace by the electric field. It
should be noted, however, that the difference between OEP and KLI
polarizabilities is on the order of 1 \%. By repeating our
calculations for different electric field strengths and on different
Cartesian grids with spacings between 0.5$a_0$ and 0.8$a_0$ and
between 65 and 129 point in each spatial direction, we estimated the
numerical accuracy of our calculations to be somewhat better than
1\%. Thus, the inaccuracies that are introduced by using the KLI
approximation are close to the numerical limits, showing once again
that, for the present systems and observables, KLI is a good
approximation to the OEP.

Compared to exchange-only LDA, the OEP yields polarizabilities which
are on average a noticeable 7\% lower. Again, this result can
readily be understood: Local exchange makes a large self-interaction
error, i.e., overestimates the electron-electron repulsion. This leads
to a density that is too
delocalized, which in turn results in too high a polarizability. 
The self-interaction error decreases with increasing electron delocalization,
and the valence electrons in sodium are nearly free electrons. Therefore,
we expect the largest self-interaction errors
and differences between local and exact-exchange polarizabilities for
the smallest clusters. This is confirmed by Table \ref{tab2}. However, the
fact that the difference falls from $\mathrm{Na}_2$ to $\mathrm{Na}_8$, but
increases somewhat from  $\mathrm{Na}_8$ to
$\mathrm{Na}_{10}$, shows that this argument seems to hold for the
overall trend, but details in the electronic configuration also seem
to matter.

Finally, comparing exact-exchange OEP to LDA for exchange and
correlation, we find the surprising result that the polarizabilities
are nearly the same. How can this be understood? The self-interaction
error for local exchange and correlation together is smaller than for
local exchange alone. So if inclusion of local correlation for the
sodium clusters would lead to perfect cancellation of the
self-interaction errors, then the fact that the polarizabilities are
practically equal would not come as a surprise. However, the local
correlation functional does not merely reduce the self-interaction
error. Correlation functionals quite generally also describe
short-range Coulomb effects that lead to stronger binding. Thus, we
expect that adding an appropriate correlation functional to the exact
exchange will bind the valence electrons more strongly, yield a more
compact density and thus lower the polarizability. Qualitatively, our
results thus indicate that the LDA overestimates the
polarizability. This conclusion, of course, is restricted to our
special situation of fixed external potential. In future work we plan
to investigate this question further by combining a newly developed,
self-interaction-free meta-GGA for correlation \cite{newmgga} with the
exact exchange functional.

\section{Summary and Conclusion}
\label{secconclusion}

In summary, we have presented a new proof for the OEP equation that
highlights the relationship between different density functionals
(e.g., Hartree-Fock and exchange-only OEP) that can be defined from a
given orbital expression for the energy. The proof explains why the
complicated OEP equation can be cast into the simple form of a
vanishing density shift. We discussed how the exact OEP can easily be
calculated from only the occupied Kohn-Sham orbitals and their
first-order shifts. The exact exchange potential was calculated for
(all-electron) closed-shell atoms and three-dimensional
(pseudopotential) sodium clusters. For the first time, not only the
total energy but also Kohn-Sham eigenvalues were accurately calculated
for three-dimensional systems from the exact OEP. We discussed
different indicators for the accuracy of an OEP solution. The KLI
approach is a rather good approximation for the ground-state
properties of atoms and small sodium clusters. The relative errors it
introduces, however, are considerably larger for the clusters than for
the atoms. We further investigated the asymptotic behavior of the
exact exchange potential of Kohn-Sham theory. For the non-spherical
clusters, it shows pronounced "ridges" which are missing in the
potentials of widely-used approximations like LDA. Finally, we
calculated the static electric dipole polarizability for several
sodium clusters and compared exact-exchange OEP to exact-exchange KLI
and to LDA. The advantages of self-interaction-free exact exchange
were demonstrated, and the need for a correlation functional to go
with exact exchange stressed.

Our method greatly facilitates the self-consistent use of orbital
functionals like the exact exchange energy in Kohn-Sham
calculations. Orbital functionals can now be used fully
self-consistently with moderate effort, and the accuracy of
approximations to the OEP, such as KLI, can be checked. Due to the
correct long-range asymptotics, functionals that use full exact
exchange appear very promising for time-dependent DFT. The great
influence that exact exchange has on the Kohn-Sham eigenvalues, which
play a prominent role in time-dependent calculations, was demonstrated
for the metal clusters. The locality of the Kohn-Sham potential is
particularly beneficial in this context and makes Kohn-Sham theory
superior\cite{graborev} to theories with orbital-dependent potentials, e.g.,
Hartree-Fock. Therefore, we believe that orbital
functionals and the OEP method will play a prominent role in future
functional development.

\begin{acknowledgments}
We are grateful to K.\ Burke for comments on the exact-exchange virial
relation, and acknowledge discussions with M.\ Levy, J.\ Tao, E.\
Sagvolden, and D.\ Kolb. We thank J.\ Krieger for asking the question
that is discussed in Appendix \ref{appendixd}. S.K.\ acknowledges
financial support by the Deutsche Forschungsgemeinschaft under an
Emmy-Noether grant, and J.P.P. by the U.S. National Science Foundation
under grant DMR 01-35678.
\end{acknowledgments}

\appendix

\section{The orbital energy density functional
$E_v^{\mathrm{orb},\alpha}[n]$} 
\label{appendixa}

Our particular definition of the orbital functional
$E_v^{\mathrm{orb},\alpha}[n]$ was chosen to avoid problems that other
rather natural-looking definitions may have. An intuitively
plausible definition, e.g, could have required the orbitals to be
orthogonal.  However, it is known that certain orbital functionals,
e.g., the SIC functional \cite{pzsic}, are minimized by non-orthogonal
orbitals. Requiring orthogonality in the definition of
$E_v^{\mathrm{orb},\alpha}[n]$ for the SIC would require non-trivial
modifications of Eq.\ (\ref{orbeq}), e.g., introducing off-diagonal
Lagrangian multipliers. On the other hand, simply dropping the
orthogonality condition and requiring the orbitals to be just the ones
minimizing the energy could lead to a ``bosonic'' solution, with all
electrons occupying the lowest orbital. Therefore, in the definition
of $E_v^{\mathrm{orb},\alpha}[n]$ the orbitals are required to be
linearly independent. They are further required to be extremizing ones
and not just the ones yielding the lowest possible energy for a given
density, because the latter criterion could be fulfilled by a "nearly
bosonic" solution, with all electrons occupying orbitals that are
extremely similar to the lowest one but with a tiny contribution added
to each one of them to make them linearly independent. Our
definition of $E_v^{\mathrm{orb},\alpha}[n]$ raises a question similar
to the questions of $v$-representability which are still a subject of
current research in Kohn-Sham theory \cite{ullrichgross}. In our case
the question is: Can we find a set of orbitals that extremizes Eq.\
(\ref{ephi}) for {\it any} given density? However, we believe that for our
present purposes this question does not pose a problem
because we are interested in the minimizing density, which we know can
be obtained from Eq. (\ref{orbeq}).

Eq.\ (\ref{orbeq}) starts from a given external potential $v(\re)$ and finds
the orbitals and density belonging to it. If instead we want the
$\varphi_i^{\mathrm{orb},\alpha}[n]$ corresponding to a given density $n(\re)$
, we must add to $v(\re)$ in Eq.\ (\ref{orbeq}) a Lagrange-multiplier $\Delta
v([n];\re)$ which enforces the constraint of Eq.\ (\ref{defn}).

\section{Proof that the minimizing densities for
$E_v^{\mathrm{orb},\alpha}[n]$ and $E_v^{\mathrm{OEP},\alpha}[n]$
agree to linear order in $\alpha$}
\label{appendixb}

In Eqs.\ (\ref{eks}) and (\ref{eorb}), we defined the functionals
$E_v^{\mathrm{OEP},\alpha}[n]$ and $E_v^{\mathrm{orb},\alpha}[n]$ by starting
from Eq.\ (\ref{ephi}) and its discussion.
Assuming that $E_v^{\mathrm{orb},\alpha}[n]$ is analytic in $\alpha$,
we consider the power series expansion
\begin{equation}
\label{eorba}
E_v^{\mathrm{orb},\alpha}[n]=E_v^{\mathrm{OEP},\alpha}[n] +\alpha
A_1[n] +\alpha^2 A_2[n] + ....
\end{equation}
The analog of this expansion for the densities is
\begin{equation}
\label{norba}
n^{\mathrm{orb},\alpha}(\re)=n^{\mathrm{OEP},\alpha}(\re)+
\alpha n_1(\re) +\alpha^2 n_2(\re)+...
\end{equation} 
We also argued in Section \ref{secdensityshift} that 
\begin{equation}
\label{a10}
A_1[n]=0.
\end{equation} 
Now rewrite the Euler equation for $n^{\mathrm{orb},\alpha}(\re)$:
\begin{eqnarray}
\lefteqn{
\left.
\frac{\delta E_v^{\mathrm{orb},\alpha}[n]}
{\delta n(\re)}\right|_{n^{\mathrm{orb},\alpha}}
=0} 
\label{step0}
\\ &=&
\left.
\frac{\delta E_v^{\mathrm{OEP},\alpha}[n]}
{\delta n(\re)}\right|_{n^{\mathrm{orb},\alpha}}
+
\left.
\alpha^2 
\frac{\delta A_2[n]}
{\delta n(\re)}\right|_{n^{\mathrm{orb},\alpha}} +...
\label{step1} 
\\ &=&
\left.
\frac{\delta E_v^{\mathrm{OEP},\alpha}[n]}
{\delta n(\re)}\right|_{n^{\mathrm{OEP},\alpha}+\alpha n_1+...}
+ \mathcal{O}(\alpha^2)
\label{step2} 
\\ &=&
\left. 
\frac{\delta E_v^{\mathrm{OEP},\alpha}[n]}
{\delta n(\re)}\right|_{n^{\mathrm{OEP},\alpha}}
\nonumber \\ &&
%\hspace{-0.7cm}
+\alpha 
\int \left. 
\frac{\delta^2 E_v^{\mathrm{OEP},\alpha}[n]}
{\delta n(\re) \delta n(\rp)}\right|_{n^{\mathrm{OEP},\alpha}}
\hspace{-0.7cm}
n_1(\rp) \, \mathrm{d}^3 r' + \mathcal{O}(\alpha^2)
\label{step3}
\end{eqnarray}
Eq.\ (\ref{step1}) follows from Eqs.\ (\ref{eorba}) and (\ref{a10}).
In Eq.\ (\ref{step2}) we use Eq.\ (\ref{norba}), and we can focus on
the terms linear in $\alpha$ because each
order of $\alpha$ must vanish separately. Eq.\ (\ref{step3}) finally
follows from functional Taylor expansion. The first term in
(\ref{step3}) is just the Euler equation for
$n^{\mathrm{OEP},\alpha}(\re)$ and therefore vanishes. So
\begin{equation}
\label{nearly}
\int \left. 
\frac{\delta^2 E_v^{\mathrm{OEP},\alpha}[n]}
{\delta n(\re) \delta n(\rp)}\right|_{n^{\mathrm{OEP},\alpha}}
%\hspace{-0.7cm}
n_1(\rp) \, \mathrm{d}^3 r'=0,
\end{equation}
and Eq.\ (\ref{nearly}) must be fulfilled for all $\alpha$. Since the
double functional derivative in Eq.\ (\ref{nearly}) depends upon
$\alpha$, while $n_1(\rp)$ does not, Eq.\ (\ref{nearly}) can only be
fulfilled by 
\begin{equation}
\label{oepagain}
n_1(\re)=0,
\end{equation}
i.e, the
minimizing densities $n^{\mathrm{orb},\alpha}(\re)$ and
$n^{\mathrm{OEP},\alpha}(\re)$ agree to linear order in $\alpha$.

We can compute $n_1(\re)$ from first-order perturbation
theory. Since, from Eq.\ (\ref{norba}),
\begin{equation}
\label{recalln}
\alpha n_1(\re)
=
n^{\mathrm{orb},\alpha}(\re)-n^{\mathrm{OEP},\alpha}(\re) 
+\mathcal{O}(\alpha^2),
\end{equation}   
$n_1(\re)$ can be computed by evaluating the density change  
that is associated with going over from the OEP-functional to the
orbital density functional and neglecting all orders in
$\alpha$ beyond the linear one. This density change can be calculated
from the corresponding change in the single-particle orbitals. To this
end, we rewrite Eq.\ (\ref{orbeq}) with the help of Eqs.\
(\ref{vxca}) and (\ref{ua}) in the form
\begin{eqnarray}
\lefteqn{  \hspace*{-8mm}
\left(
 -\frac{\hbar^2}{2m}\nabla^2 + v(\re) + v_\mathrm{H}(\re) +
 \alpha v_\mathrm{xc}(\re)
\right) \varphi_i^{\mathrm{orb},\alpha}(\re) 
} 
\nonumber \\ && \hspace*{-8mm}
+\alpha \left[
 u_{\mathrm{xc}i}(\re)- v_\mathrm{xc}(\re)
\right] \varphi_i^{\mathrm{orb},\alpha}(\re)
= 
\varepsilon_i^{\mathrm{orb},\alpha}\varphi_i^{\mathrm{orb},\alpha}(\re).
\label{perteq}
\end{eqnarray}
Clearly, if we neglect the term in brackets in the second line of Eq.\
(\ref{perteq}), we are looking at the Kohn-Sham system. Thus, we
can calculate 
the shift that the orbitals undergo when we go over from the Kohn-Sham
to the orbital density functional by treating 
$\alpha [ u_{\mathrm{xc}i}(\re)- v_\mathrm{xc}(\re)]$
as a perturbation on the Kohn-Sham system. In non-degenerate
first-order perturbation theory (see Section
\ref{secdensityshift} for remarks on the degenerate case), the shifted
orbital is 
\begin{equation}
\label{phi1sto}
\varphi_{i}^{\mathrm{pert},\alpha}(\re)=
\varphi_{i}^\alpha(\re) +
\sum_{\stackrel{\scriptstyle j=1}{j\neq i}}^{\infty}
\frac{\langle \varphi_j^\alpha 
  |\alpha \left[ u_{\mathrm{xc}i} - v_\mathrm{xc} \right] |
  \varphi_i^\alpha \rangle 
}
{\varepsilon_{i}-\varepsilon_{j} }
\varphi_{j}^\alpha(\re), \nonumber 
\end{equation}
where $\varphi_i^{\alpha}(\re)$ and $\varepsilon_i$ are the Kohn-Sham
orbitals and eigenvalues corresponding to the minimizing density
$n^{\mathrm{OEP},\alpha}(\re)$. The important fact to be noted here is
that the perturbation is linear in $\alpha$. Thus, higher orders of
perturbation theory will also be of higher order in $\alpha$. Since we
want to calculate the change in density only to first order in
$\alpha$, only the orbital shift in first-order perturbation theory
needs to be taken into account. We insert Eq.\ (\ref{phi1sto}) into
Eq.\ (\ref{recalln}),
\begin{eqnarray}
\alpha n_1(\re)
&=&
\sum_{i=1}^N \varphi_{i}^{\mathrm{pert},\alpha}(\re) 
             {\varphi_{i}^{\mathrm{pert},\alpha}}^*(\re)
-
\sum_{i=1}^N \varphi_{i}^{\alpha}(\re) 
             {\varphi_{i}^{\alpha}}^*(\re)
\nonumber \\
&=& 
\alpha \sum_{i=1}^N \varphi_{i}^{\alpha}(\re) 
             {\varphi_{i,1}^{\alpha}}^*(\re) + c.c. + \mathcal{O}(\alpha^2),
\end{eqnarray}
where $\varphi_{i,1}^{\alpha}(\re)$ is defined as the second term on the
right-hand side of Eq.\ (\ref{phi1sto}) divided by
$\alpha$. Obviously, $\varphi_{i,1}^{\alpha=1}(\re)=-\psi_i(\re)$, and 
thus Eq.\ (\ref{oepagain}) is exactly Eq.\ (\ref{oep}).

\section{Numerical aspects of the iterative construction}
\label{appendixc}

In this Appendix we summarize some of the more technical details of the
iterative solution of the OEP equation. 

We first discuss the convergence of the iteration based on Eq.\
(\ref{oepexpl2}) for the exact-exchange-only OEP. The test system here
is the Be atom. The Kohn-Sham equations were solved on a logarithmic
radial grid. Table \ref{tab3} illustrates how many iterations going
back and forth between Eq.\ (\ref{kseq}) and Eq.\ (\ref{oepexpl2}) are
necessary for a given number of
($\psi_{i\sigma}^*,v_{\mathrm{xc}\sigma}$)-cycles to converge the
total energy and eigenvalues to the OEP values with an accuracy of 0.1
mhar. The starting point is the KLI potential. The total energy is
correct after one iteration in all cases (the difference is small to
begin with), but also the eigenvalues which are noticeably different
in KLI and OEP converge quickly: With just 5
($\psi_{i\sigma}^*,v_{\mathrm{xc}\sigma}$)-cycles per iteration, it
takes only 5 iterations to reach the exact OEP. We also note that for
the second and later iteration steps, the self-consistent solution of
the Kohn-Sham equations
%during the OEP iterations 
proceeds much faster than the first self-consistent solution that is
needed to get the starting approximation. This has two reasons.
First, we need to recalculate the $u_{\mathrm{xc}i\sigma}$ only once
at the beginning of each new
($\psi_{i\sigma}^*,v_{\mathrm{xc}\sigma}$)-cycle and not at each step
in the self-consistent solution of the Kohn-Sham equations (since
$v_\mathrm{xc}$ is kept fixed during the latter). Second, the orbitals
and shifts from the previous iteration are already a good starting
guess for the next one. Therefore, the whole scheme converges very
rapidly.
\begin{table}[b]
\caption{Number of iterations necessary to converge all eigenvalues of
the Be atom to their exact OEP values $E=-14.5724$,
$\varepsilon_{1}=-4.1257$, $\varepsilon_{2}=-0.3092$, for a given
number of ($\psi_{i\sigma}^*,v_{\mathrm{xc}\sigma}$)-cycles. 
The total energy is correct after one iteration. Starting
point is the KLI approximation ($E=-14.5723$,
$\varepsilon_{1}=-4.1668$,
$\varepsilon_{2}=-0.3089$, all energies in hartree). 
Note that 
convergence is achieved with very moderate effort.} 
\label{tab3}
\begin{ruledtabular}
\begin{tabular}{cc|cc}
($\psi_{i\sigma}^*,v_{\mathrm{xc}\sigma}$)-cycles & iterations &
($\psi_{i\sigma}^*,v_{\mathrm{xc}\sigma}$)-cycles & iterations  \\ \hline
1 & 24 &  5 & 5 \\
2 & 10 & 10 & 3 \\
3 &  9 & 20 & 3
\end{tabular}
\end{ruledtabular}
\end{table}

However, Eq.\ (\ref{oepexpl2})
requires 
dividing by the density, and this is a disadvantage if it is to be used for
finite systems. The first term under the
sum is unproblematic. It is asymptotically dominated by the highest 
occupied orbital which also
dominates the density. Thus, even if the density in the denominator
has decayed to the 
extent that its numerical representation becomes highly inaccurate,
the same inaccuracies show up in the numerator, and the correct result
is still obtained. This cancellation does not occur in the
second or gradient term, and the resulting inaccuracies can make the scheme
unstable. For spherical systems, the problem can easily 
be solved. In the asymptotic region, the KLI potential is already correct,
therefore the second term need not be evaluated numerically there. And
the asymptotic region for a spherical system can easily be
identified by going out from the system's center and monitoring when
the density starts to be dominated by the highest occupied
orbital. In this way, calculations for spherical atoms can be done simply.  

But the accuracy of the method depends rather sensitively on the
cut-off point: With a cut-off too close to the system's center, too
much of the second term is lost; with a cut-off that is too far away,
the iteration becomes unstable. In the spherical, effectively
one-dimensional calculations, the numerical parameters can be chosen
such that finding an appropriate cut-off radius is possible. Our real
interest, however, is in non-spherical systems of possibly complex
shape. We found that for three-dimensional (3D) systems, the technical
difficulties are more severe. The coarser grids that computational
necessity forces on us lead to larger inaccuracies in the evaluation
of the derivatives that appear in the second term and allow for less
fine tuning of the cut-off point. Furthermore, the asymptotic region
will start at different distances in different directions and must be
mapped out more carefully. 

As a first test case for a 3D
implementation of our algorithm, we chose the jellium cluster
$\mathrm{Na}_8$. It is a good test case because it is spherical, i.e.,
the OEP can be obtained from the established direct solution of the
integral equation for comparison \cite{engel2}. But unlike the
diverging atomic nuclear potential, jellium $\mathrm{Na}_8$ can also
be represented on the 3D Cartesian grid which we use for a realistic,
pseudopotential-based description of metal clusters. The Seitz radius
($r_s=[3/(4\pi n)]^{1/3}$) of the jellium background density was
$r_s=3.93 \, a_0$, corresponding 
to a sphere of radius $7.86 \, a_0$. We first calculated the KLI
approximation and compared a spherical, one-dimensional calculation on
a logarithmic grid to a 3D calculation (Cartesian grid, 65 points in
each direction, spacing $0.7 a_0$, damped gradient iteration with
finite-difference multigrid relaxation \cite{diss}) that does not
exploit the spherical symmetry. Both yield identical energies and
eigenvalues to within 0.1 mhar. This confirms the accuracy of our 3D
code. We then constructed the OEP iteratively on the same grid for
different cut-off radii between $4\,a_0$ and $19\, a_0$. Only with
cut-offs close to $6\, a_0$ was the iteration unproblematic and
converged to the correct total energy E=-0.3735 har. However, the
lowest eigenvalue was still 0.4 mhar above its correct value 0.2081
har, i.e., halfway between the OEP and KLI values. Obtaining the OEP
via Eq.\ (\ref{oepexpl2}) for the real, non-spherical clusters turned
out to be even more complicated. The situation can be improved with
finer grids, but using Eq.\ (\ref{oepexpl2}) in 3D calculations is
rather tedious. 

The iteration based on Eq.\ (\ref{siter}) does not suffer from these
problems. The convergence is slower than for the first method, but
the computational effort is still very moderate: For the test system
jellium $\mathrm{Na}_8$, the total energy and all eigenvalues converge
within 5 iterations if 10 cycles are used. With Eq.\ (\ref{siter}),
the 3D calculation now also yields exactly the same energy and eigenvalues as
the spherical, direct solution of the OEP integral equation
\cite{talman,engel2}. During the iterations, the relative error in the
exchange virial relation falls by two orders of magnitude compared to
the KLI approximation, and $S_\mathrm{max}$ decreases to $2\times
10^{-6}$. Further iterations reduce both quantities further.
Thus, the simple iteration very accurately yields the exact OEP also
for 3D systems.

The parameter $c$ used in the iteration was chosen by trial and
error. If $c$ is chosen too large, the iteration diverges, and with $c$
too small, it does not converge as quickly as possible. In all cases we
studied, finding appropriate values for $c$ was unproblematic and
there was a range of suitable values. We used $c$'s between 3 har$\,
a_0^3$ and 0.75 har$\, a_0^3$ for the atoms, and $c$=30 har$\, a_0^3$
for all the cluster calculations. Thus, $c$ was similar for similar
systems, i.e., $c$'s of about 1 were appropriate for the atoms with
their singular nuclear potential, and larger $c$'s could be used for
the non-singular pseudopotential calculations. 

These values can be understood on the basis of Eq.\
(\ref{nearlythere}). For $\mathrm{Na}_4$, e.g.,
$-v_{\mathrm{x}\sigma}(\re)/(2 n_\sigma(\re))$ takes values between 30
har$\, a_0^3$ and 40 har$\, a_0^3$ in the interior region of the
cluster. A yet simpler and charming argument can be made based on the
bulk limit. For a homogenous system, 
$-v_\mathrm{x}/n = (9 \pi/4)^{(1/3)} r_s^2 \, \mathrm{har} \, a_0 
\approx 1.92 \, r_s^2$ har$\, a_0$. For bulk
sodium, $r_s=3.93 \, a_0$, so $-v_\mathrm{x}/n \approx$ 30 har$\, a_0^3$, and
for atoms, the average density is about $r_s=1 \, a_0$, so
$-v_\mathrm{x}/n$ is about 2 har$\, a_0^3$. These estimates agree
nicely with the values we found by trial-and-error. Finally, we want
to remark that the iteration could also be used with some suitably
chosen $c(\re)$, but for our present purposes a constant was
satisfactory.

\section{On the difference between KLI and OEP orbital shifts}
\label{appendixd}

Looking at Fig.\ \ref{fig1}, one might ask the following question: The
KLI and OEP orbitals are rather similar. Therefore, in Eq.\
(\ref{psipde}) the operator on the left-hand side and all quantities
on the right-hand side should be similar, too. But then, also the
orbital shifts that are calculated from Eq.\ (\ref{psipde}) should be
similar. So how can it be that we observe the pronounced difference
between the KLI and OEP orbital shift $\psi_1(\re)$ shown in the
lowest part of Fig.\ \ref{fig1}?

The answer to this equestion is that although the Kohn-Sham orbitals
obtained from the KLI potential and the OEP are similar, the
respective right-hand sides in Eq.\ (\ref{psipde}) nevertheless show a
qualitative difference. The higher Kohn-Sham orbital $\varphi_2(\re)$
has a nodal surface in the x-y plane, i.e., its orbital density
vanishes there. Therefore, in this plane the KLI potential is just
given by
$u_{\mathrm{x}1}(\re)-(\bar{v}_{\mathrm{x}1}-\bar{u}_{\mathrm{x}1})$.
If this is inserted for $v_\mathrm{x}(\re)$ in the right-hand side of
Eq.\ (\ref{psipde}), the right-hand side vanishes. But for the OEP,
there is an additional contribution to $v_\mathrm{x}(\re)$ from the
term of Eq.\ (\ref{oepexpl2}) that involves the derivatives of the
Kohn-Sham orbitals and their orbital shifts, and this term makes a
non-vanishing contribution in the x-y plane. The qualitative
difference between OEP and KLI orbital shift is thus a direct
consequence of a qualitative difference of the respective right-hand
sides in Eq.\ (\ref{psipde}).

%Figures

%Tables

\end{document}